\documentclass[preprint,authoryear,12pt]{elsarticle}

\usepackage{mathrsfs}
\usepackage[T1]{fontenc} 
\usepackage[USenglish]{babel} 
\usepackage{graphicx} 
\usepackage{makeidx} 
\usepackage{remreset} 
\usepackage[nouppercase]{scrpage2} 
\usepackage{amsmath} 
\usepackage{amssymb} 
\usepackage[
thmmarks, 
amsmath, 
hyperref 
]{ntheorem} 
\usepackage{bm} 
\usepackage{bbm} 
\usepackage{enumitem} 
\usepackage[hang]{subfigure} 
\usepackage{wrapfig} 
\usepackage{tabularx} 
\usepackage{multirow}
\usepackage{dcolumn} 
\usepackage{booktabs} 
\usepackage{listings} 
\usepackage{psfrag} 

\usepackage[table]{xcolor}
\usepackage{appendix}
\usepackage{caption}
\usepackage{capt-of}

\usepackage{algorithmic,algorithm}
\numberwithin{algorithm}{section}

\newcommand{\cali}{{\cal I}}
\newcommand{\calj}{{\cal J}}
\newcommand{\calk}{{\cal K}}

\usepackage[
setpagesize=false, 
pdfborder={0 0 0}, 
pdfpagemode=UseOutlines, 
]{hyperref} 
\makeatletter
\renewcommand{\p@enumii}[1]{\theenumi(#1)}
\makeatother

\theoremstyle{break} 
\theoremheaderfont{\bfseries}
\theorembodyfont{\itshape}
\theoremseparator{}
\newtheorem{definition}{Definition}[section] 

\newtheorem{example}[definition]{Example}
\theoremstyle{nonumberbreak} 
\theoremsymbol{$\Box$}

\newcommand*{\IR}{\mathbb{R}}

\newcommand{\btheta}{\mbox{\boldmath $\theta$}}

\makeatletter
\renewcommand*\env@matrix[1][*\c@MaxMatrixCols c]{%
  \hskip -\arraycolsep
  \let\@ifnextchar\new@ifnextchar
  \array{#1}}
\makeatother

\journal{-}

\begin{document}

\begin{frontmatter}


\title{Is there significant time-variation in multivariate copulas?}


	\author{Jakob St{\"o}ber and Ulf Schepsmeier }

\address{Center for Mathematical
Sciences, Technische Universit\"at M\"unchen, Germany. Corresponding author email: \texttt{stoeber@ma.tum.de}}

\begin{abstract}
We demonstrate how the uncertainty of parameter point estimates can be assessed in a maximum likelihood framework in order to prevent overfitting and erroneous detection of time-inhomogeneity.
The class of models we consider are regular vine (R-vine) copula models, for which we describe a new algorithm for the exact computation of the score function and observed information. R-vine copulas constitute a flexible class of dependence models which are constructed hierarchically from bivariate copulas as building blocks only, and our algorithm exploits the hierarchical nature for subsequent computation of log-likelihood derivatives. Results obtained using the proposed methods are discussed in the context of the asymptotic efficiency of different estimation methods for R-vine based models. In a substantial application to a dataset of exchange rates, we obtain clear indications for time-inhomogeneous dependence between some currency pairs.
\end{abstract}		


\end{frontmatter}

\section{Introduction}


The last years have seen the rise of dependence models for complex multivariate data and the renewed experience during the recent (2008) financial crisis that well-accepted economic paradigms may loose their validity from one day to another.
This has created significant interest in the study of time-homogeneity of multivariate dependence (see e.g. \citet{pelletier2006}, \citet{manner2011}, \citet{cholette2009} and \citet{garcia2011}).
However, while it is a standard exercise in multivariate statistics to compute the uncertainty incorporated in parameter point estimates for classes like the multivariate normal distribution, this is often not possible for the more complex models which are required to accurately capture the dependence in real-world multivariate data.
A particularly successful class of dependence models are regular vine (R-vine) copula models which have been introduced in a subsequent series of papers by \citet{Joe2},  \citet{Bedford_Cooke2001,Bedford_Cooke}, \citet{kurowicka:cooke:2006},  \citet{Aas_Czado} and \citet{DissmannBrechmannCzadoKurowicka2011}. They have been applied to model dependence in various areas including agricultural science and electricity loads \citet{smith:min:czado:almeida}, exchange rates (\citet{CzadoSchepsmeierMin2011}, \citet{stoeber2011c}), order books and headache data \citet{panagiotelis2012}. 
In general, an R-vine model is constituted by a set of bivariate copulas corresponding to conditional distributions determined by a sequence of linked trees. Following \citet{Bedford_Cooke2001}, the trees $(T_1,\ldots,T_{d-1})$ forming an R-vine tree sequence $\mathcal{V}$ are required to fulfill the following properties: 
\begin{enumerate}
	\item $T_1$ is a tree with nodes $N_1=\{1,\ldots,d\}$ and edges $E_1$.
	\item For $i\geq 2$, $T_i$ is a tree with nodes $N_i = E_{i-1}$ and edges $E_i$.
	\item If two nodes in $T_{i+1}$ are joint by an edge, the corresponding edges in $T_i$ must share a common node {\it (proximity condition)}.
\end{enumerate}
An 8-dimensional example is given in Figure	\ref{fig:pre-rvine}.
Following the notation of \citet{Czado} with a set of bivariate copula densities\\ $\mathcal{B} = \left\{c_{j(e),k(e)|D(e)} \vert e \in E_i, 1 \leq i \leq d-1 \right\}$ corresponding to edges $j(e),k(e)|D(e)$ in $E_i$, for $1\leq i\leq d-1$ the density of a $d$-dimensional R-vine distribution with marginal densities $f_k$,  $1\leq k\leq d$ is given by
\begin{equation}
\begin{split}
	&f_{1,\ldots,d}(x_1,\ldots,x_d)\\
	& \ \ =\prod_{i=1}^d f_i(x_i)\prod_{i=1}^{d-1}\prod_{e\in E_i}c_{j(e),k(e)|D(e)}(F_{j(e)|D(e)}(x_{j(e)}|\boldsymbol{x}_{D(e)}),F_{k(e)|D(e)}(x_{k(e)}|\boldsymbol{x}_{D(e)})).
	\label{eq:density}
\end{split}
\end{equation} 
Here, $\boldsymbol{x}_{D(e)}$ is the subvector of $\boldsymbol{x}$ determined by the set of indices in $D(e)$, which is called {\it conditioning set} while the indices $j(e)$ and $k(e)$ form the {\it conditioned set}.
If all marginal densities are uniform, the corresponding distribution is called an R-vine copula. 
Often it will be convenient to split the specification of bivariate parametric copula families in $\mathcal{B}$ from the respective parameters which are stored in an additional vector $\btheta$. In this case, an R-vine copula is specified in terms of $R\mathcal{V}=(\mathcal{V}, \mathcal{B}, \btheta)$. 

Despite the wide range of applications of R-vine copula based models in practice, there is a surprising scarcity in the literature considering the uncertainty in point estimates of the copula parameters. 
It is well known that maximum likelihood estimates $\hat{\btheta}_n$ will be strongly consistent and asymptotically normal under regularity conditions on the bivariate building blocks, i.e. 
\begin{equation}
\sqrt{n}\ \cali(\boldsymbol \theta)^{1/2}\Big(\hat{\boldsymbol \theta}_n - \boldsymbol \theta\Big) \stackrel{d}{\longrightarrow} N(0,\boldsymbol{Id}_p)\ \text{as}\ n\rightarrow\infty.
\label{eq:asymptotic_MLE}
\end{equation}
Here, $\boldsymbol \theta$ is the true $p$-dimensional parameter vector, $n$ is the number of (i.i.d.) observations,  $\boldsymbol{Id}_p$ the $p\times p$ identity matrix, and 
{
\small
\begin{equation}
\label{fisher_information}
\cali(\boldsymbol \theta)=-\mathbbm{E}_{\boldsymbol \theta}\Bigg[\Big( \frac{\partial^2}{\partial \theta_i \partial \theta_j } l(\boldsymbol \theta\vert \boldsymbol{X}) \Big)_{i,j=1,\ldots,p}\Bigg] = \mathbbm{E}_{\boldsymbol \theta}\Bigg[\Big( \frac{\partial}{\partial \theta_i } l(\boldsymbol \theta \vert \boldsymbol{X}) \cdot \frac{\partial}{\partial \theta_j } l(\boldsymbol \theta\vert \boldsymbol{X}) \Big)_{i,j=1,\ldots,p}\Bigg],
\end{equation}
}\noindent
denotes the \emph{Fisher Information Matrix} with $l(\boldsymbol{\theta}\vert \boldsymbol{x})$ being the log-likelihood of parameter $\btheta$ for one observation $\boldsymbol{x}$. Given the fact that full maximum likelihood inference is numerically difficult when non-uniform marginal distributions are involved, also two-step procedures have been developed. 
In particular, \citet{joe1996} employ the probability integral transform using parametric marginal distributions which are fitted in a first step to obtain uniform (copula) data on which the copula is estimated using ML in a second step. 
As an alternative, \citet{genest1995} proposed to use non-parametric rank transformations in the first step. 
While these methods are computationally more tractable, they are asymptotically less efficient. Following \citet{haff2010}, we can decompose the asymptotic covariance matrix for the estimates of dependence parameters $\boldsymbol{\theta}$ in a marginal part and a dependence part:
\[
	\boldsymbol V^{\boldsymbol \theta, two-step} = \boldsymbol V^{dependence} + \boldsymbol V^{margins},
\]
where $\boldsymbol V^{dependence,ML}=\cali(\boldsymbol \theta)^{-1}$ for ML estimation and the second part is zero only when there is no uncertainty about the margins. Further, also the estimation of copula parameters can be performed in a tree by tree fashion. Denote the log-likelihood arising from parameters in tree $i$ by
{
\small
$$ l_i(\btheta_i\vert\boldsymbol{x}):=\sum_{e\in E_i} \log\left( c_{j(e),k(e)|D(e)}(F_{j(e)|D(e)}(x_{j(e)}|\boldsymbol{x}_{D(e)}),F_{k(e)|D(e)}(x_{k(e)}|\boldsymbol{x}_{D(e)})\vert \theta_e) \right),$$
}\noindent
where the set of components of $\btheta_i$ is $\left\{ \theta_e \vert e \in E_i \right\}$. In particular $\btheta_1$ is estimated by maximizing $l_1(\btheta_1 \vert \boldsymbol{x})$ and the obtained estimates $\hat{\btheta}_1$ are used to calculate the arguments of the copula functions in $l_2(\btheta_2 \vert \boldsymbol{x})$. Now, we maximize $l_2(\btheta_2 \vert \boldsymbol{x})$ in $\btheta_2$ to obtain $\hat{\btheta}_2$ and proceed until all parameters are estimated (for an algorithm see \citet{stoeber2011c}). This implies that $l_i$ also implicitly depends on the parameters $\boldsymbol{\theta}_j$ for $j<i$ through its arguments, i.e. $l_i(\btheta_i \vert \boldsymbol{x})=l_i(\btheta_i, \hat{\btheta}_{i-1}, \ldots,\hat{\btheta}_1 \vert \boldsymbol{x})$.
The parameter estimates obtained from this sequential procedure have asymptotical covariance
\begin{equation}
\boldsymbol V^{dependence,seq.} =\boldsymbol{\calj}_{\boldsymbol \theta}^{-1}\boldsymbol{\calk}_{\boldsymbol \theta}\left(\boldsymbol{\calj}_{\boldsymbol \theta}^{-1}\right)^T,
\label{seqerrors}
\end{equation}
where $\boldsymbol{\calj}_{\boldsymbol \theta}$ involves second derivatives of the R-vine log-likelihood function and $\boldsymbol{\calk}_{\boldsymbol \theta}$ involves elements of the score function, see \citet{haff2010}. To be more precise, $\boldsymbol{\calk}_{\boldsymbol \theta}$ and $\boldsymbol{\calj}_{\boldsymbol \theta}$ are defined as
\begin{equation}
	\boldsymbol{\calk}_{\boldsymbol\theta} = \begin{pmatrix}
	\boldsymbol{\calk}_{\boldsymbol{\theta},1,1} \\
	\vdots & \ddots \\
	\boldsymbol{0}^T & \cdots & \boldsymbol{\calk}_{\boldsymbol{\theta},d-2,d-2} \\
	\boldsymbol{0}^T & \cdots & \boldsymbol{0}^T & \boldsymbol{\calk}_{\boldsymbol{\theta},d-1,d-1} \\
	\end{pmatrix},
	\label{eq:matrixK}
\end{equation}
\begin{equation}
	\boldsymbol{\calj}_{\boldsymbol \theta} = \begin{pmatrix}
	\boldsymbol{\calj}_{\boldsymbol{\theta},1,1} \\
	\vdots & \ddots \\
	\boldsymbol{\calj}_{\boldsymbol{\theta},d-2,1} & \cdots & \boldsymbol{\calj}_{\boldsymbol{\theta},d-2,d-2} \\
	\boldsymbol{\calj}_{\boldsymbol{\theta},d-1,1} & \cdots & \boldsymbol{\calj}_{\boldsymbol{\theta},d-1,d-2} & \boldsymbol{\calj}_{\boldsymbol{\theta},d-1,d-1} \\
	\end{pmatrix},
	\label{eq:matrixJ}
\end{equation}
with $\boldsymbol{\calk}_{\boldsymbol{\theta},i,j} = E\left(\left(\frac{\partial l_i(\boldsymbol{\theta}_i,\ldots,\boldsymbol{\theta}_1|\boldsymbol{X})}{\partial \boldsymbol{\theta}_i}\right)\left(\frac{\partial l_j(\boldsymbol{\theta}_j,\ldots,\boldsymbol{\theta}_1|\boldsymbol{X})}{\partial \boldsymbol{\theta}_j}\right)^T\right)$ and \\ $\boldsymbol{\calj}_{\boldsymbol{\theta},i,j} = -E\left(\frac{\partial^2 l_i(\boldsymbol{\theta}_i,\ldots,\boldsymbol{\theta}_1|\boldsymbol{X})}{\partial\boldsymbol{\theta}_i\partial\boldsymbol{\theta}_j}\right), i,j=1,\ldots,d-1$.

While this asymptotic theory is well known, it is almost never applied in practice since the estimation of the asymptotic covariance matrix will involve the Hessian matrix, i.e. the second derivatives of the R-vine likelihood function. 
For these derivatives, no analytical expressions have been available creating a gap between theoretical knowledge about estimation errors and practical applicability which we will fill in this paper.

The remainder is structured as follows: Section \ref{sec:likelihood} shows how the (log-) likelihood of a general R-vine model can be calculated efficiently. Based on the illustrated algorithm, Sections \ref{sec:firstDerivative} and \ref{sec:secondDerivative} consider the computation of the first and second derivatives of the R-vine copula log-likelihood with respect to the parameters or the score function and observed information, respectively. In Section \ref{sec:application} we set to answering the headline question of this paper the developed algorithms to a dataset of exchange rates and Section \ref{sec:discussion} concludes by wrapping up our results.

\section{Computation of the R-vine likelihood}\label{sec:likelihood}
In order to calculate the (log-) likelihood function of an R-vine model, we must develop an algorithmic way to evaluate the copula terms in the decomposition (\ref{eq:density}) with respect to the appropriate arguments. 
Here, the R-vine structure with the proximity condition implies that for each edge $e$ the term
\begin{equation}
\begin{split}
	F(x_{j(e)}|\boldsymbol{x}_{D(e)}) &= \frac{\partial C_{j(e),j'(e)|D(e)\setminus j'(e)}(F(x_{j(e)}|\boldsymbol{x}_{D(e) \setminus j^{\prime}(e)}),F(x_{j^{\prime}(e)}|\boldsymbol{x}_{D(e)\setminus j^{\prime}(e)}))}{\partial F(x_{j^{\prime}(e)}|\boldsymbol{x}_{D(e)\setminus j^{\prime}(e)})} \\
	&=: h_{j(e),j'(e)|D(e)\setminus j'(e)}(F(x_{j(e)}|\boldsymbol{x}_{D(e)\setminus j^{\prime}(e)}),F(x_{j^{\prime}(e)}|\boldsymbol{x}_{D(e)\setminus j^{\prime}(e)})),
\end{split}
\label{eq:hfunction}
\end{equation}
and similarly $F(x_{k(e)}|\boldsymbol{x}_{D(e)})$ can be computed. This means, that there is an index $j'(e) \in D(e)$, such that the copula $C_{j(e),j'(e)|D(e)\setminus j'(e)}$ is in $\mathcal{B}$. For this expression, the assumption that the copula $ C_{j(e),j'(e)|D(e)\setminus j'(e)}$ does not depend on the values $\boldsymbol{x}_{D(e)\setminus j^{\prime}(e)}$ is made. This is called {\it simplifying assumption}, as it will simplify further computations as well as model selection. For which classes of multivariate distributions this assumption is or is not applicable is discussed in \citet{Stoeber2012}. 
To further ease notation, we will assume that all copulas under investigation are symmetric in their arguments, such that we do not have to differentiate between $h_{j(e),j'(e)|D(e)\setminus j'(e)}$ and $h_{j'(e),j(e)|D(e)\setminus j'(e)}$. While this is valid for most common parametric copula families, we can easily drop this assumption later.

\begin{figure}
	\centering
		\includegraphics[width=0.80\textwidth]{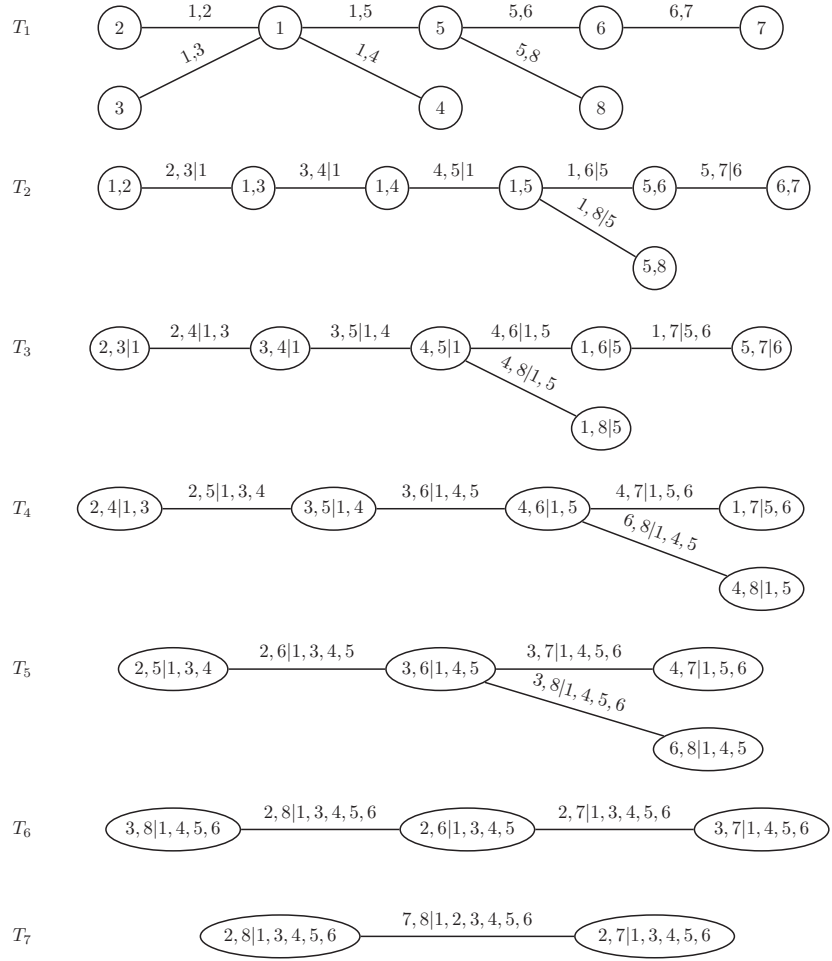}
	\caption{An R-vine tree sequence in 8 dimensions with edge indices corresponding to the pair-copulas in an R-vine copula model.}
	\label{fig:pre-rvine}
\end{figure}

In order to perform computations for a general R-vine copula model, it is convenient to use a matrix notation which has been introduced by \citet{MoralesNapolesCookeKurowicka2009}, \citet{Dissmann2010} and \citet{DissmannBrechmannCzadoKurowicka2011}. It stores the edges of an R-vine tree sequence in the following way:
Consider for example the R-vine in Figure \ref{fig:pre-rvine} which can be described in matrix notation as follows:

{\scriptsize\begin{equation}
 M=\begin{pmatrix}
 m_{1,1}\\
 m_{2,1} & m_{2,2} \\
 m_{3,1} & m_{3,2} & m_{3,3} \\
 m_{4,1} & m_{4,2} & m_{4,3} & m_{4,4} \\
 m_{5,1} & m_{5,2} & m_{5,3} & m_{5,4} & m_{5,5} \\
 m_{6,1} & m_{6,2} & m_{6,3} & m_{6,4} & m_{6,5} & m_{6,6} \\
m_{7,1} & m_{7,2} & m_{7,3} & m_{7,4} & m_{7,5} & m_{7,6} & m_{7,7} \\
m_{8,1} & m_{8,2} & m_{8,3} & m_{8,4} & m_{8,5} & m_{8,6} & m_{8,7} & m_{8,8} \\
 \end{pmatrix}=\begin{pmatrix}
 8\\
 7 & 7 \\
 2 & 2 & 6 \\
 3 & 3 & 2 & 5 \\
 6 & 4 & 3 & 2 & 4 \\
 4 & 1 & 4 & 3 & 2 & 3 \\
 1 & 5 & 1 & 4 & 3 & 2 & 2 \\
 5 & 6 & 5 & 1 & 1 & 1 & 1 & 1\\
 \end{pmatrix}.
 \label{eq:matrixM}
 \end{equation}}\noindent
For example the edge index $74|156$ is stored by $m_{2,2}, m_{5,2}$ given $m_{6,2}, m_{7,2}$ and $m_{8,2}$. 
Accordingly, we can store the copula families $\mathcal{B}$ and the corresponding parameters $\btheta$. 
{\footnotesize 
 \begin{equation*}
 \btheta=\begin{pmatrix}
 \ldots \\
 \ldots  & \theta_{4, m_{6,5}\vert m_{7,5},m_{8,5}} \\
 \ldots  & \theta_{4, m_{7,5}\vert m_{8,5}} & \theta_{3, m_{7,6}\vert m_{8,6}} \\
 \ldots  & \theta_{4, m_{8,5}} & \theta_{3, m_{8,6}} & \theta_{2, m_{8,7}} & \ \\
 \end{pmatrix} = \begin{pmatrix}
 \ldots \\
 \ldots &\theta_{2,4\vert 1,3} \\
 \ldots  & \theta_{3,4\vert 1} & \theta_{2,3\vert 1} \\
 \ldots & \theta_{1,4} & \theta_{1,3} & \theta_{1,2} & \ \\
 \end{pmatrix}
 \end{equation*} }

As an illustration for how the R-vine matrix is derived from the pictured tree sequence  in Figure \ref{fig:pre-rvine} and vice versa, let us consider the third column of the matrix. Here we have $6$ on the diagonal, and $2$ as a second entry. The set of remaining entries below $2$ is $\{3, 4, 1, 5 \}$. This corresponds to the edge $2,6\vert 1,3,4,5$ in $T_5$ of Figure \ref{fig:pre-rvine}.
Similarly, the edge $3,6\vert 1,4,5$ corresponds to the third entry $3$ in the third column, $1,6\vert 4,5$ to the fourth entry, etc.
Note that the diagonal of $M$ is sorted in descending order which can always be achieved by reordering the node labels. From now on, we will assume that all matrices are "normalized" in this way as this allows to simplify notation. Therefore we have $m_{i,i}=d-i+1$. For further shortening and clarity of index labels, we will illustrate the notations in the example of an $8$-dimensional vine.

Applying the matrix notation, the (log-)likelihood of an R-vine model is computed by first storing all required conditional distribution functions evaluated at a $d$-dimensional vector of observations $(u_1,\ldots, u_d)$ in two matrices. In particular, we calculate

{\footnotesize\begin{equation}
\label{Vdirect}
V^{direct}=\begin{pmatrix}
\ldots & & & &  \\
\ldots & F(u_4\vert u_{m_{6,5}}, u_{m_{7,5}},u_{m_{8,5}}) & &  &  \\
\ldots &  F(u_4 \vert u_{m_{7,5}}, u_{m_{8,5}}) & F(u_3 \vert u_{m_{7,6}},u_{m_{8,6}}) & &  \\
\ldots & F(u_4\vert u_{m_{8,5}}) & F(u_3\vert u_{m_{8,6}}) &  F(u_2\vert u_{m_{8,7}})   \\
\ldots  & u_4 & u_3 & u_2 & u_1 & \  \\
\end{pmatrix}
\end{equation}
\begin{equation}
\label{Vindirect}
V^{indirect}=\begin{pmatrix}
\ldots & & & & & \\
\ldots & F(u_{m_{6,5}}\vert u_{m_{7,5}},u_{m_{8,5}},u_4) & &  & &\ \\
\ldots &  F(u_{m_{7,5}}\vert u_{m_{8,5}}, u_4) & F(u_{m_{7,6}} \vert u_{m_{8,6}},u_3) & & &\ \\
\ldots & F(u_{m_{8,5}}\vert u_4) & F( u_{m_{8,6}}\vert u_3) &  F(u_{m_{8,7}}\vert u_2)  &\ \\
\ldots & u_{m_{8,5}} & u_{m_{8,6}} & u_{m_{8,7}} &  \ \\
\end{pmatrix}.
\end{equation}}

Note that, for each pair-copula term in (\ref{eq:density}), the corresponding terms of $V^{direct}$ and $V^{indirect}$ can be easily determined. When being able to evaluate
$$c_{4,m_{6,5} \vert  m_{7,5}, m_{8,5}}(F(u_4 \vert u_{m_{7,5}}, u_{m_{8,5}}),F(u_{m_{6,5}} \vert u_{m_{7,5}}, u_{m_{8,5}}))$$ we do also obtain 
{\small\begin{equation*}
\begin{split}
F(u_4\vert u_{m_{6,5}}, &u_{m_{7,5}},u_{m_{8,5}})= \\
&=(\partial_1 C)_{4,m_{6,5}\vert  m_{7,5}, m_{8,5}}(F(u_4 \vert u_{m_{7,5}}, u_{m_{8,5}}),F(u_{m_{6,5}} \vert u_{m_{7,5}}, u_{m_{8,5}})),\\
F(u_{m_{6,5}} \vert u_4, &u_{m_{7,5}},u_{m_{8,5}})= \\
&=(\partial_2 C)_{4,m_{6,5}\vert  m_{7,5}, m_{8,5}}(F(u_4 \vert u_{m_{7,5}}, u_{m_{8,5}}),F(u_{m_{6,5}} \vert u_{m_{7,5}}, u_{m_{8,5}})),
\end{split}
\end{equation*}}\noindent
where $(\partial_1C)$ and $(\partial_2 C)$ denote partial derivatives with respect to the first and second argument, respectively, c.f. Equation (\ref{eq:hfunction}).
With all such conditional distribution functions being available, the copula terms in (\ref{eq:density}) corresponding to the next tree $T_4$ can be evaluated.
This sequential calculation is performed in Algorithm \ref{alg:LogLikelihood},  which was developed in \citet{Dissmann2010} and \citet{DissmannBrechmannCzadoKurowicka2011}. Following the notation in (\ref{eq:hfunction}), we write $h( \cdot , \cdot \vert \mathcal{B}^{k,i}, \theta^{k,i})$ for the conditional distribution function corresponding to a parametric family $\mathcal{B}^{k,i}$ with parameter $\theta^{k,i}$, where $\mathcal{B}^{k,i}$ and $\theta^{k,i}$ denote the $(k,i)$th element of the matrices $\mathcal{B}$ and $\btheta$, respectively. Exempli gratia, 
\begin{equation*}
\begin{split}
F(u_4 \vert u_2, u_3, u_1) &= h(F(u_4 \vert u_3, u_1), F(u_2 \vert u_3, u_1) \vert \mathcal{B}_{4,2\vert 3,1}, \theta_{4,2\vert 3,1}) \\
&= h(v_{6,5}^{direct},v_{6,6}^{indirect}\vert \mathcal{B}^{6,5}, \theta^{6,5}).
\end{split}
\end{equation*}
The only question which is left to solve for the computation of the (log-) likelihood is whether the arguments in each step (i.e. $F(u_4 \vert u_3, u_1), F(u_2 \vert u_3, u_1)$ in the example) have to be picked from the matrix $V^{direct}$ or $V^{indirect}$. For this, we exploit the descending order of the diagonal of $M$. From the structure of $V^{direct}$, we see that the first argument of the copula term with family $\mathcal{B}^{k,i}$ and parameter $\theta^{k,i}$ is stored as the $(k,i)$th element $v^\text{direct}_{k,i}$ of $V^{direct}$.
To locate the second entry, let us denote $\tilde{M} = (\tilde{m}_{k,i} |i = 1, \ldots, d; k = i, \ldots, d)$, where $\tilde{m}_{k,i} := \max \{ m_{k,i}, \ldots, m_{d,i} \}$ for all $i = 1, \ldots, d$ and $k = i, \ldots, d$. The second argument, which is $F(u_{m_{k,i}} \vert m_{i,i}, m_{k+1,i}, \ldots, m_{d,i})$ must be in column $(d-\tilde{m}_{k,i}+1)$ of $V^{direct}$ or $V^{indirect}$ by the ordering of variables. If $\tilde{m}_{k,i} =  m_{k,i}$, the conditioning variable $u_{m_{k,i}}$ has the biggest index and thus the entry we are looking for must be in $V^{direct}$. Similarly, if $\tilde{m}_{k,i} > m_{k,i}$, the variable with the biggest index is in the conditioning set and we must choose from $V^{indirect}$.

\begin{example}[Selection of arguments for $c_{2,4|1,3}$]
As an example for how this procedure selects the correct arguments for copula terms in the regular vine let us consider the copula  $c_{2,4|1,3}$ in our example distribution.  The corresponding parameter $\theta_{2,4|1,3}$ is stored as $\theta^{6,5}$, thus we are in the case where $i=5$ and $k=6$. Since $\tilde{m}_{6,5} = \max\{m_{6,5}, m_{7,5}, m_{8,5}\} = \max\{2,3,1\} = 3$ and $\tilde{m}_{6,5} = 3 > 2 = m_{6,5}$ we select as second argument the entry $v_{k,(d-\tilde{m}_{k,i}+1)}^{indirect} = v_{6,6}^{indirect}=F(u_2 \vert u_1 , u_3)$. Together with $v^{direct}_{6,5}=F(u_4\vert u_1, u_3)$ which we have already selected, this is the required argument.
\end{example}

The corresponding algorithm to compute the log-likelihood of an R-vine specification for a single observation $\boldsymbol{u}=(u_1,\ldots,u_d)$ is given in Algorithm \ref{alg:LogLikelihood}.

\begin{algorithm}[ht]
\caption{Log-likelihood of an R-vine specification.}
\label{alg:LogLikelihood}
\begin{algorithmic}[1]
	\REQUIRE $d$-dimensional R-vine specification in matrix form, i.e., $M$, $\mathcal{B}$, $\btheta$, set of observations $(u_1,\ldots, u_d)$.
	\STATE Set $L = 0$.\label{alg:LogLikelihood:F1}
	\STATE Let $V^{\text{direct}} = (v^\text{direct}_{k,i} | i = 1, \ldots, d; k=i,\ldots,d)$.
	\STATE Let $V^{\text{indirect}} = (v^\text{indirect}_{k,i} | i = 1, \ldots, d; k=i,\ldots,d)$.
	\STATE Set $(v^\text{direct}_{d,1}, v^\text{direct}_{d,2}, \ldots, v^\text{direct}_{d,d})
				= (u_d, u_{d-1}, \ldots u_1)$.
	\STATE Let $\tilde{M} = (\tilde{m}_{k,i} | i = 1, \ldots, d; k = i, \ldots, d)$ where $\tilde{m}_{k,i} = \max \{ m_{k,i}, \ldots, m_{d,i} \}$
	for all $i = 1, \ldots, d$ and $k = i, \ldots, d$.
	\FOR[Iteration over the columns of $M$]{$i = d-1, \ldots, 1$}
		\FOR[Iteration over the rows of $M$]{$k = d, \ldots, i+1$}
			\STATE Set $z_1 = v^\text{direct}_{k,i}$ \label{alg:LogLikelihood:z1}
			\IF{$\tilde{m}_{k,i} =  m_{k,i}$} \label{alg:LogLikelihood:IF}
				\STATE Set $z_2 = v^\text{direct}_{k,(d-\tilde{m}_{k,i}+1)}$.
				 \label{alg:LogLikelihood:IF:1}
			\ELSE
				\STATE Set $z_2 = v^\text{indirect}_{k,(d-\tilde{m}_{k,i}+1)}$.
				 \label{alg:LogLikelihood:IF:2}
			\ENDIF
			\STATE Set $L = L + c(z_1,z_2 | \mathcal{B}^{k,i},  \theta^{k,i})$.
			\label{alg:LogLikelihood:LL}			
			\STATE Set $v^\text{direct}_{k-1,i} =  h(z_1,z_2 | \mathcal{B}^{k,i}, \theta^{k,i})$ and $v^\text{indirect}_{k-1,i} = h(z_2,z_1 | \mathcal{B}^{k,i},  \theta^{k,i})$.
			\label{alg:LogLikelihood:h}	
		\ENDFOR
	\ENDFOR
	\RETURN L
\end{algorithmic}
\end{algorithm}

\section{Computation of the score function}
\label{sec:firstDerivative}

In this section we develop an algorithm to calculate the derivatives of the R-vine log-likelihood with respect to copula parameters and thus the score function of the model. Throughout the remainder, we will assume that all occurring copula densities are continuously differentiable with respect to their arguments and parameters. Further, we assume that the copula parameters are all in $\IR$, the extension to two or higher dimensional parameter spaces is straightforward but makes the notation unnecessarily complex.

To determine the log-likelihood derivatives, we will again exploit the hierarchical structure of the R-vine copula model and proceed similarly as for the likelihood calculation. The first challenge which we must overcome to develop an algorithm for the score function is to determine which of the copula terms in Expression (\ref{eq:density}) depend on which parameter directly or indirectly through one of their arguments. Following the steps of the log-likelihood computation and exploiting the structure of the R-vine structure matrix $M$, this is decided in Algorithm \ref{alg_1}. 

\begin{algorithm}[ht]
\caption{Determine copula terms which depend on a specific parameter.}
\label{alg_1}
The input of the algorithm is a $d$-dimensional R-vine matrix $M$ with elements $(m_{l,j})_{l,j=1,\dots,d}$ and the row number $k$ and column number $i$ corresponding to the position of the parameter of interest in the corresponding parameter matrix $\btheta$. 
The output will be a matrix $C$ (with elements $(c_{l,j})_{l,j=1,\dots,d}$)  of zeros and ones, a one indicating that the copula term corresponding to this position in the matrix will depend on the parameter under consideration.
\begin{algorithmic}[1]
\STATE Set $g:=(m_{i,i}, m_{k,i}, m_{k+1,i}, \dots, m_{d,i})$
\STATE Set $c_{l,j}:=0 \ \ l,j=1,\dots, d$
\FOR{$a = i, \dots, 1$}
	\FOR{$b = k,\dots,a+1$}
		\STATE Set $h:=(m_{a,a}, m_{b,a}, m_{b+1,a},\dots, m_{d,a})$
		\IF{$\#(g \cap h) == \# g$}
			\STATE Set $c_{b,a}:=1$
		\ENDIF
	\ENDFOR
\ENDFOR
\RETURN C
\end{algorithmic}
\end{algorithm}

Knowing how a specific copula term depends on a given parameter, we can proceed with calculating the corresponding derivatives. 
Before we explain the derivatives in detail let us start with an example where two of the three possible cases of dependence on a given parameter are illustrated.

\begin{example}[3-dim]
Let $x_1\sim F_1, x_2\sim F_2, x_3\sim F_3$ and $u_1=F(x_1), u_2=F(x_2)$, $u_3=F(x_3)$, then the joint density can be decomposed as
\begin{equation*}
\begin{split}
	f_{123}(x_1,x_2,x_3) &= f_1(x_1)f_2(x_2)f_3(x_3)\cdot c_{1,2}(u_1,u_2\vert\theta_{1,2})\cdot c_{2,3}(u_2,u_3\vert\theta_{2,3}) \\
	&\cdot c_{1,3|2}(h_{1,2}(u_2,u_1\vert\theta_{1,2}),h_{2,3}(u_3,u_2\vert \theta_{2,3}) \vert \theta_{1,3|2})
\end{split}
\end{equation*}
The first derivatives of $\ln f_{123}$ with respect to the copula parameters are
\begin{equation*}
\begin{split}
	&\frac{\partial (\ln f_{123}(x_1,x_2,x_3))}{\partial \theta_{1,2}} = \frac{ \partial_{\theta_{1,2}} c_{1,2}(u_1,u_2\vert\theta_{1,2}) }{ c_{1,2}(u_1,u_2\vert\theta_{1,2}) } \\
	&\ \ +  \frac{ \partial_1 c_{1,3|2}(h_{1,2}(u_2,u_1\vert\theta_{1,2}),h_{2,3}(u_3,u_2\vert\theta_{2,3})\vert\theta_{1,3|2}) }{c_{1,3|2}(h_{1,2}(u_2,u_1\vert\theta_{1,2}),h_{2,3}(u_3,u_2\vert\theta_{2,3})\vert\theta_{1,3|2}) }\cdot \partial_{\theta_{1,2}} h_{1,2}(u_1,u_2\vert\theta_{1,2}) \\
	&\frac{\partial (\ln f_{123}(x_1,x_2,x_3))}{\partial \theta_{2,3}} = \frac{ \partial_{\theta_{2,3}} c_{2,3}(u_2,u_3\vert\theta_{2,3}) }{ c_{2,3}(u_2,u_3\vert\theta_{2,3}) } \\
	&\ \ +  \frac{ \partial_2 c_{1,3|2}(h_{1,2}(u_2,u_1\vert\theta_{1,2}),h_{2,3}(u_3,u_2\vert\theta_{2,3})\vert\theta_{1,3|2}) }{c_{1,3|2}(h_{1,2}(u_2,u_1\vert\theta_{1,2}),h_{2,3}(u_3,u_2\vert\theta_{2,3})\vert\theta_{1,3|2}) }\cdot \partial_{\theta_{2,3}} h_{2,3}(u_2,u_3\vert\theta_{2,3}) \\
	&\frac{\partial (\ln f_{123}(x_1,x_2,x_3))}{\partial \theta_{1,3|2}} = \frac{ \partial_{\theta_{1,3|2}} c_{1,3|2}(h_{1,2}(u_2,u_1\vert\theta_{1,2}),h_{2,3}(u_3,u_2\vert\theta_{2,3})\vert\theta_{1,3|2}) }{c_{1,3|2}(h_{1,2}(u_2,u_1\vert\theta_{1,2}),h_{2,3}(u_3,u_2\vert\theta_{2,3})\vert\theta_{1,3|2}) }.
\end{split}
\end{equation*}
\label{example1}
\end{example}

The first case which occurs in our example is that the copula densities $c_{1,2}$ and $c_{2,3}$ depend on their respective parameters directly. For a general term involving a copula $c_{U,V\vert \textbf{Z}}$ with parameter $\theta$, 
{
\small 
\begin{equation}
\begin{split}
\frac{\partial}{\partial \theta} \ln\left( c_{U,V\vert \textbf{Z}}\left( F_{U\vert\textbf{Z}}( u \vert \textbf{z}), F_{V\vert\textbf{Z}}( v \vert \textbf{z})\vert \theta \right) \right) &= 
\frac{ \frac{\partial}{\partial \theta}  \left(c_{U,V\vert \textbf{Z}}\left( F_{U\vert\textbf{Z}}( u \vert \textbf{z}), F_{V\vert\textbf{Z}}( v \vert \textbf{z})\vert \theta \right) \right) }{ c_{U,V\vert \textbf{Z}}\left( F_{U\vert\textbf{Z}}( u \vert \textbf{z}), F_{V\vert\textbf{Z}}( v \vert \textbf{z})\vert \theta \right)} \\
&= \frac{ \partial_\theta  c_{U,V\vert \textbf{Z}}\left( F_{U\vert\textbf{Z}}( u \vert \textbf{z}), F_{V\vert\textbf{Z}}( v \vert \textbf{z})\vert \theta \right) }{ c_{U,V\vert \textbf{Z}}\left( F_{U\vert\textbf{Z}}( u \vert \textbf{z}), F_{V\vert\textbf{Z}}( v \vert \textbf{z})\vert \theta \right)}.
\end{split}
\label{eq:copuladeriv3}
\end{equation}}\noindent
Further, like for $c_{1,3\vert 2}$, a $c_{U,V\vert \textbf{Z}}$ term can depend on a parameter $\theta$ through one of its arguments, say $F_{U\vert \textbf{Z}}(u \vert \textbf{z}, \theta)$:
{\small 
\begin{equation}
\begin{split}
\label{eq:copuladeriv1}
\frac{\partial}{\partial \theta} \ln\big( c_{U,V\vert \textbf{Z}}\big( &F_{U\vert\textbf{Z}}( u \vert \textbf{z}, \theta ), F_{V\vert\textbf{Z}}( v \vert \textbf{z}) \big) \big) = \\
&=\ \ \frac {\frac{\partial c_{U,V\vert \textbf{Z}}\left( F_{U\vert\textbf{Z}}( u \vert \textbf{z}, \theta ), F_{V\vert\textbf{Z}}( v \vert \textbf{z} ) \right) }{\partial F_{U\vert\textbf{Z}}(u \vert \textbf{z}, \theta)} }{  c_{U,V\vert \textbf{Z}}\left( F_{U\vert\textbf{Z}}( u \vert \textbf{z}, \theta ), F_{V\vert\textbf{Z}}( v \vert \textbf{z} ) \right) } \cdot \frac{\partial}{\partial \theta} F_{U\vert\textbf{Z}}(u \vert \textbf{z}, \theta) \\
&= \frac {\partial_1 c_{U,V\vert \textbf{Z}}\left( F_{U\vert\textbf{Z}}( u \vert \textbf{z}, \theta ), F_{V\vert\textbf{Z}}( v \vert \textbf{z} ) \right)}{  c_{U,V\vert \textbf{Z}}\left( F_{U\vert\textbf{Z}}( u \vert \textbf{z}, \theta ), F_{V\vert\textbf{Z}}( v \vert \textbf{z} ) \right) } \cdot \frac{\partial}{\partial \theta} F_{U\vert\textbf{Z}}(u \vert \textbf{z}, \theta).
\end{split}
\end{equation}}\noindent
Finally, in dimension $d\geq 4$, both arguments of a $c_{U,V\vert \textbf{Z}}$ copula term can depend on a parameter $\theta$. In this case,
{\small \begin{equation}
\begin{split}
 \frac{\partial}{\partial \theta} \ln\big( c_{U,V\vert \textbf{Z}}\big(& F_{U\vert\textbf{Z}}( u \vert \textbf{z}, \theta ), F_{V\vert\textbf{Z}}( v \vert \textbf{z}, \theta ) \big) \big)\\
 = & \frac {\partial_1 c_{U,V\vert \textbf{Z}} \left( F_{U\vert\textbf{Z}}( u \vert \textbf{z}, \theta ), F_{V\vert\textbf{Z}}( v \vert \textbf{z}, \theta ) \right)}{  c_{U,V\vert \textbf{Z}}\left( F_{U\vert\textbf{Z}}( u \vert \textbf{z}, \theta ), F_{V\vert\textbf{Z}}( v \vert \textbf{z}, \theta ) \right) }  \cdot  \frac{\partial}{\partial \theta} F_{U\vert\textbf{Z}}(u \vert \textbf{z}, \theta) \  \\
 + &  \frac {\partial_2 c_{U,V\vert \textbf{Z}} \left( F_{U\vert\textbf{Z}}( u \vert \textbf{z}, \theta ), F_{V\vert\textbf{Z}}( v \vert \textbf{z}, \theta ) \right)}{  c_{U,V\vert \textbf{Z}}\left( F_{U\vert\textbf{Z}}( u \vert \textbf{z}, \theta ), F_{V\vert\textbf{Z}}( v \vert \textbf{z}, \theta ) \right) } \cdot \frac{\partial}{\partial \theta} F_{V\vert\textbf{Z}}(v \vert \textbf{z}, \theta).
\end{split}
\label{eq:copuladeriv2}
\end{equation}}\noindent
We see that the derivatives of copula terms corresponding to tree $T_i$ in the vine will involve derivatives of conditional distribution functions which are determined by tree $T_{i-1}$. Thus, it will be convenient to store their derivatives in matrices $S1^{direct,\theta}$ and $S1^{indirect,\theta}$ related to the matrices $V^{direct}$ and $V^{indirect}$ which have been determined during the calculation of the log-likelihood together with the terms 
\begin{equation*}
\ln\Big(c_{j(e),k(e)|D(e)}(F_{j(e)|D(e)}(x_{j(e)}|\boldsymbol{x}_{D(e)}),F_{k(e)|D(e)}(x_{k(e)}|\boldsymbol{x}_{D(e)}))\Big)=:\varrho_{j(e),k(e)|D(e)},
\end{equation*}
for each edge $e$ in the R-vine $\mathcal{V}$ , which can also be stored in a matrix $V^{values}$:
\begin{equation}
\label{pcc_rvine_alg_matrices_3}
V^{values}=\begin{pmatrix}
\ldots & & & & \\
\ldots & \varrho_{4, m_{6,5} \vert  m_{7,5},  m_{8,5} }& &  &  \\
\ldots &  \varrho_{4, m_{7,5} \vert m_{8,5}} &\varrho_{3, m_{7,6} \vert m_{8,6}} & &  \\
\ldots & \varrho_{4, m_{8,5}}& \varrho_{3, m_{8,6}} &  \varrho_{2,1}   \\
\ldots  & &  &  &  \ \\
\end{pmatrix}
\end{equation}
In particular, we will determine the following matrices:
{\footnotesize\begin{equation}
\label{pcc_rvine_alg_matrices_1_tilde}
S1^{direct,\theta}=\begin{pmatrix}
\ldots &  \\
\ldots & \frac{\partial}{\partial \theta}F(u_4\vert u_{m_{6,5}},u_{m_{7,5}},u_{m_{8,5}}) & &  & \\
\ldots &  \frac{\partial}{\partial \theta}F(u_4 \vert u_{m_{7,5}}, u_{m_{8,5}}) & \frac{\partial}{\partial \theta}F(u_3 \vert u_{m_{7,6}},u_{m_{8,6}}) & &\\
\ldots & \frac{\partial}{\partial \theta} F(u_4\vert u_{m_{8,5}}) & \frac{\partial}{\partial \theta} F(u_3\vert u_{m_{8,6}}) &  \frac{\partial}{\partial \theta} F(u_2\vert u_1)  \\
\ldots  & &  &    \\
\end{pmatrix}
\end{equation}}
{\footnotesize\begin{equation}
\label{pcc_rvine_alg_matrices_2_tilde}
S1^{indirect,\theta}=\begin{pmatrix}
\ldots & & \\
\ldots & \frac{\partial}{\partial \theta} F(u_{m_{6,5}}\vert u_{m_{7,5}},u_{m_{8,5}},u_4) &\\
\ldots & \frac{\partial}{\partial \theta} F(u_{m_{7,5}}\vert u_{m_{8,5}}, u_4) & \frac{\partial}{\partial \theta} F(u_{m_{7,6}} \vert u_{m_{8,6}},u_3) &  \\
\ldots &\frac{\partial}{\partial \theta}  F(u_{m_{8,5}}\vert u_4) & \frac{\partial}{\partial \theta} F( u_{m_{8,6}}\vert u_3) & \frac{\partial}{\partial \theta}  F(u_1\vert u_2) &    \\
\ldots  &  \\
\end{pmatrix}
\end{equation}}
{\small \begin{equation}
\label{pcc_rvine_alg_matrices_3_tilde}
S1^{values,\theta}=\begin{pmatrix}
\ldots & & & & \\
\ldots & \frac{\partial}{\partial \theta} \varrho_{4, m_{6,5} \vert  m_{7,5},  m_{8,5} }& &  &  \\
\ldots &  \frac{\partial}{\partial \theta} \varrho_{4, m_{7,5} \vert m_{8,5}} & \frac{\partial}{\partial \theta} \varrho_{3, m_{7,6} \vert m_{8,6}} & &  \\
\ldots & \frac{\partial}{\partial \theta} \varrho_{4, m_{8,5}}&  \frac{\partial}{\partial \theta} \varrho_{3, m_{8,6}} &   \frac{\partial}{\partial \theta} \varrho_{2,1}   \\
\ldots  & &  &  &  \ \\
\end{pmatrix}.
\end{equation}}\noindent
Here, the terms in $S1^{direct,\theta}$ and $S1^{indirect,\theta}$ can be determined by differentiating (\ref{eq:hfunction}) similarly as we did for the copula terms in (\ref{eq:copuladeriv3}) - (\ref{eq:copuladeriv2}). For instance, we have
\begin{equation*}
\begin{split}
 \frac{\partial}{\partial \theta} F_{U\vert V,\textbf{Z} }(u \vert v,\textbf{z}, \theta) &= \frac{\partial}{\partial \theta}  \left (h_{U\vert V, \textbf{Z}}\left( F_{U\vert\textbf{Z}}( u \vert \textbf{z}, \theta ), F_{V\vert\textbf{Z}}( v \vert \textbf{z}) \right) \right) \\ &=  \partial_1 h_{U\vert V,\textbf{Z} }\left( F_{U\vert\textbf{Z}}( u \vert \textbf{z}, \theta ), F_{V\vert\textbf{Z}}( v \vert \textbf{z}) \right) \cdot \frac{\partial}{\partial \theta}  F_{U\vert\textbf{Z}}( u \vert \textbf{z}, \theta )  \\ 
 &= c_{U\vert V,\textbf{Z} }\left( F_{U\vert\textbf{Z}}( u \vert \textbf{z}, \theta ), F_{V\vert\textbf{Z}}( v \vert \textbf{z}) \right) \cdot \frac{\partial}{\partial \theta}  F_{U\vert\textbf{Z}}( u \vert \textbf{z}, \theta )
 \end{split}
 \end{equation*}
 and
 \begin{equation*}
\begin{split}
 \frac{\partial}{\partial \theta} h_{U\vert V,\textbf{Z} }\big( F_{U\vert\textbf{Z}}( &u \vert \textbf{z} ), F_{V\vert\textbf{Z}}( v \vert \textbf{z}, \theta) \big)= \\ 
 &= \partial_2 h_{U\vert V,\textbf{Z} }\left( F_{U\vert\textbf{Z}}( u \vert \textbf{z} ), F_{V\vert\textbf{Z}}( v \vert \textbf{z}, \theta) \right) \cdot \frac{\partial}{\partial \theta}  F_{V\vert\textbf{Z}}( v \vert \textbf{z}, \theta ).
 \end{split}
 \end{equation*}
The complete calculations required to obtain the derivative of the log-likelihood with respect to one copula parameter $\theta$ are performed in Algorithm \ref{alg_2}.

\begin{center}
  \captionsetup{style=ruled,type=algorithm,skip=0pt}
  \makeatletter
    \fst@algorithm\@fs@pre
  \makeatother
  \caption{Log-likelihood derivative with respect to the parameter $\theta^{\tilde k, \tilde i}$. }
  \label{alg_2}
  \makeatletter
    \@fs@mid
  \makeatother

\begin{flushleft}
The input of the algorithm is a $d$-dimensional R-vine matrix $M$ with maximum matrix $\tilde M$ and parameter matrix $\btheta$, and a matrix $C$ determined using Algorithm \ref{alg_1} for a parameter $\theta^{\tilde{k},\tilde{i}}$ positioned at row $\tilde k$ and $\tilde i$ in the R-vine parameter matrix $\boldsymbol{\theta}$. 
Further, we assume the matrices $V^{direct}$, $V^{indirect}$ and $V^{values}$ corresponding to one observation from the R-vine copula distribution, which have been determined during the calculation of the log-likelihood, to be given. The output will be the value of the first derivative of the copula log-likelihood for the given observation with respect to the parameter $\theta^{\tilde k, \tilde i}$.
\end{flushleft}

\begin{algorithmic}[1]
\STATE Set $z_1= v^{direct}_{\tilde k, \tilde i}$
\STATE Set $s1^{direct}_{k,i}:=0$, $s1^{indirect}_{k,i}:=0$, $s1^{values}_{k,i}:=0$, $i=1,\ldots,d; k=i,\ldots,d$
\IF{$m_{\tilde k, \tilde i}== \tilde{ m}_{\tilde k, \tilde i}$}
	\STATE Set $z_2=v^{direct}_{\tilde k, d - \tilde{ m}_{\tilde k, \tilde i} + 1}$
\ELSE
	\STATE Set $z_2=v^{indirect}_{\tilde k, d - \tilde{ m}_{\tilde k, \tilde i} + 1}$
\ENDIF
\STATE Set $s1^{direct}_{\tilde k -1,\tilde i} = \partial_{\theta_{\tilde k,\tilde i}} h(z_1,z_2 \vert \mathcal{B}^{\tilde{k},\tilde{i}}, \theta^{\tilde{k},\tilde{i}})$
\STATE Set $s1^{indirect}_{\tilde k -1,\tilde i} = \partial_{\theta_{\tilde k,\tilde i}} h(z_2,z_1 \vert \mathcal{B}^{\tilde{k},\tilde{i}}, \theta^{\tilde{k},\tilde{i}})$
\STATE Set $s1^{values}_{\tilde{k},\tilde{i}} = \frac{\partial_{\theta_{\tilde k,\tilde i}} c(z_1,z_2 \vert \mathcal{B}^{\tilde{k},\tilde{i}}, \theta^{\tilde{k},\tilde{i}})}{exp(v^{values}_{\tilde k, \tilde i})}$
\FOR{$i=\tilde i, \dots, 1$}
	\FOR{$k=\tilde k -1, \dots, i+1$}
		\IF{$c_{k,i}==1$}
			\STATE Set $z_1= v^{direct}_{k, i}$, $\tilde{z}_1= s1^{direct}_{k, i}$
			\IF{$m_{ k, i}==\tilde{m}_{k,i}$}
				\STATE Set ${z}_2={v}^{direct}_{ k, d - \tilde{m}_{k,i} + 1}$, $\tilde{z}_2=s1^{direct}_{ k, d - \tilde{m}_{k,i} + 1}$
			\ELSE
				\STATE Set ${z}_2={v}^{indirect}_{ k, d - \tilde{m}_{k,i} + 1}$, $\tilde{z}_2=s1^{indirect}_{ k, d - \tilde{m}_{k,i} + 1}$
			\ENDIF
			\IF{$c_{k+1,i}==1$}
				\STATE Set $s1^{values}_{k,i}= s1^{values}_{k,i} + \frac{\partial_1 c( z_1, z_2 \vert \mathcal{B}^{k,i}, \theta^{k,i} )}{v^{values}_{k, i}} \cdot \tilde{z}_1$
				\STATE Set $s1^{direct}_{k-1,i}= s1^{direct}_{k-1,i} + \partial_1 h( z_1, z_2 \vert \mathcal{B}^{k,i}, \theta^{k,i} ) \cdot \tilde{z}_1$
				\STATE Set $s1^{indirect}_{k-1,i}= s1^{indirect}_{k-1,i} + \partial_2 h( z_2, z_1 \vert \mathcal{B}^{k,i}, \theta^{k,i} ) \cdot \tilde{z}_1$
			\ENDIF
			\IF{$c_{k+1,d-m+1}== 1$}
				\STATE Set $s1^{values}_{k,i}= s1^{values}_{k,i} + \frac{\partial_2 c( z_1, z_2 \vert \mathcal{B}^{k,i}, \theta^{k,i} )}{exp(v^{values}_{k, i})} \cdot \tilde{z}_2$
				\STATE Set $s1^{direct}_{k-1,i}= s1^{direct}_{k-1,i} + \partial_2 h( z_1, z_2 \vert \mathcal{B}^{k,i}, \theta^{k,i} ) \cdot \tilde{z}_2$
				\STATE Set $s1^{indirect}_{k-1,i}= s1^{indirect}_{k-1,i} + \partial_1 h( z_2, z_1 \vert \mathcal{B}^{k,i}, \theta^{k,i} ) \cdot \tilde{z}_2$
			\ENDIF
		\ENDIF
	\ENDFOR
\ENDFOR
\RETURN $\sum_{k,i = 1, \dots,d} s1^{values}_{k,i} $
\end{algorithmic}
\makeatletter
    \@fs@post
  \makeatother
\end{center}
In particular, this algorithm allows to replace finite-differences based numerical maximization of R-vine likelihood functions with maximization based on the analytical gradient. In a numerical comparison study across different R-vine models in 5-8 dimensions this resulted in a decrease in computation time by a factor of 4-8.

\section{Computation of the observed information}
\label{sec:secondDerivative}

Based on the calculation of the score function performed in the previous section, we will present an algorithm to determine the Hessian matrix corresponding to the R-vine log-likelihood function in this section. 

Again, we employ a convenient matrix notation. Considering a derivative with respect to bivariate copula parameters $\theta$ and $\gamma$ associated with the vine, it is clear that the expressions for the derivatives of the log-densities in this case will contain second derivatives of the occurring h-functions. Thus, our algorithm will determine the following matrices:

{\small\begin{equation}
\label{pcc_rvine_alg_matrices_1_bar}
\begin{split}
&S2^{direct,\theta,\gamma}= \\
& \begin{pmatrix}
\ldots & & & & & \\
\ldots & \frac{\partial}{\partial \theta \partial \gamma}F(u_4\vert u_{m_{6,5}},u_{m_{7,5}},u_{m_{8,5}}) & &  & &\ \\
\ldots &  \frac{\partial}{\partial \theta\partial \gamma}F(u_4 \vert u_{m_{7,5}}, u_{m_{8,5}}) & \frac{\partial}{\partial \theta\partial \gamma}F(u_3 \vert u_{m_{7,6}},u_{m_{8,6}}) & & &\ \\
\ldots & \frac{\partial}{\partial \theta\partial \gamma} F(u_4\vert u_{m_{8,5}}) & \frac{\partial}{\partial \theta\partial \gamma} F(u_3\vert u_{m_{8,6}}) &  \frac{\partial}{\partial \theta\partial \gamma} F(u_2\vert u_1)  &\ \\
\ldots  & &  &  &  \\
\end{pmatrix}
\end{split}
\end{equation}}
{\small\begin{equation}
\begin{split}
\label{pcc_rvine_alg_matrices_2_bar}
&S2^{indirect,\theta,\gamma}= \\
&\begin{pmatrix}
\ldots & & & & & \\
\ldots & \frac{\partial}{\partial \theta\partial \gamma} F(u_{m_{6,5}}\vert u_{m_{7,5}},u_{m_{8,5}},u_4) & &  &  \\
\ldots & \frac{\partial}{\partial \theta\partial \gamma} F(u_{m_{7,5}}\vert u_{m_{8,5}}, u_4) & \frac{\partial}{\partial \theta\partial \gamma} F(u_{m_{7,6}} \vert u_{m_{8,6}},u_3) & &  \\
\ldots &\frac{\partial}{\partial \theta\partial \gamma}  F(u_{m_{8,5}}\vert u_4) & \frac{\partial}{\partial \theta\partial \gamma} F( u_{m_{8,6}}\vert u_3) & \frac{\partial}{\partial \theta\partial \gamma}  F(u_1\vert u_2)   \\
\ldots  & &  &     \\
\end{pmatrix}
\end{split}
\end{equation}}
{\small\begin{equation}
\label{pcc_rvine_alg_matrices_3_bar}
S2^{values,\theta,\gamma}= 
\begin{pmatrix}
\ldots & & & & \\
\ldots & \frac{\partial}{\partial \theta\partial \gamma} \varrho_{4, m_{6,5} \vert  m_{7,5},  m_{8,5} }& &  &  \\
\ldots & \frac{\partial}{\partial \theta\partial \gamma} \varrho_{4, m_{7,5} \vert m_{8,5}} & \frac{\partial}{\partial \theta\partial \gamma} \varrho_{3, m_{7,6} \vert m_{8,6}} & &  \\
\ldots & \frac{\partial}{\partial \theta\partial \gamma} \varrho_{4, m_{8,5}}&  \frac{\partial}{\partial \theta\partial \gamma} \varrho_{3, m_{8,6}} &   \frac{\partial}{\partial \theta\partial \gamma} \varrho_{2,1}   \\
\ldots  & &  &  &  \ \\
\end{pmatrix}.
\end{equation}}

Since not all entries in (\ref{Vdirect}), (\ref{Vindirect}) and (\ref{pcc_rvine_alg_matrices_3}) depend on both $\theta$ and $\gamma$, not all entries in (\ref{pcc_rvine_alg_matrices_1_bar}) - (\ref{pcc_rvine_alg_matrices_3_bar}) will be non-zero and required in the algorithm. Employing Algorithm \ref{alg_1} to obtain matrices $C^\theta$ and $C^\gamma$ corresponding to the parameters $\theta$ and $\gamma$, respectively, we see that the second derivatives of all elements where the corresponding matrix entry of either $C^\theta$ or $C^\gamma$ is zero clearly vanish.

To derive an algorithm similar to Algorithm \ref{alg_2} which recursively determines all terms of the second derivatives of the log-likelihood with respect to parameters $\theta, \gamma$, we need to distinguish 7 basic cases of dependence on the two parameters which can occur for a term $c_{U,V\vert \textbf{Z}}\left(F_{U \vert \textbf{Z}}(u\vert \textbf{z}), F_{V \vert \textbf{Z}}(v \vert \textbf{z})\right)$.

\begin{center}
\begin{tabular}{c | c | c | c  }
cases & dependence on & $\theta$ & $\gamma$ \\
\hline
case 1 & \multirow{7}{*}{through} & $F_{U \vert \textbf{Z}}$ & $F_{V \vert \textbf{Z}}$ \\
case 2 & & $F_{U \vert \textbf{Z}}, F_{V \vert \textbf{Z}}$ & $F_{V \vert \textbf{Z}}$ \\
case 3 & & $F_{U \vert \textbf{Z}}, F_{V \vert \textbf{Z}}$ & $F_{U \vert \textbf{Z}}$, $F_{V \vert \textbf{Z}}$ \\
case 4 & & $F_{U \vert \textbf{Z}}$ & $c_{U,V\vert \textbf{Z}}$ \\
case 5 & & $F_{U \vert \textbf{Z}},F_{V \vert \textbf{Z}}$ & $c_{U,V\vert \textbf{Z}}$ \\
case 6 & & $F_{U \vert \textbf{Z}}$ & $F_{U \vert \textbf{Z}}$ \\
case 7 & & $c_{U,V\vert \textbf{Z}}$ & $c_{U,V\vert \textbf{Z}}$ \\
\hline
\end{tabular}
 \end{center}
Here, case 7 is relevant only for derivatives where $\theta=\gamma$, since we assume that all bivariate copulas occurring in the vine density have one parameter in $\IR$.
Because of symmetry in the parameters, all other possible combinations are already included in these cases, we only have to exchange $\theta$ and $\gamma$. A more detailed description of the occurring derivatives is given in \ref{appendix:algorithms}.

As before, the terms in $S2^{direct,\theta,\gamma}$ and $S2^{indirect,\theta,\gamma}$ can be determined by differentiating (\ref{eq:hfunction}) similarly as we did for the copula terms in (\ref{second_derivative_first_equation}) - (\ref{second_derivative_last_equation}) of \ref{appendix:algorithms}. Thus, the second derivatives can again be calculated recursively (see Algorithm \ref{alg_3}).

Combining Algorithm \ref{alg_3} and numerical integration techniques\footnote{We use the adaptive integration routines supplied by Steven G. Johnson and Balasubramanian Narasimhan
in the \textit{cubature} package available on CRAN which are based on \citet{genz1980} and \citet{berntsen1991}.} we can also calculate the Fisher information (see Equation (\ref{fisher_information})) matrix of R-vine copula models and determine asymptotical standard errors for ML estimates.

\begin{example}[3-dim. Gaussian and Student t-copula vine models]
Let us consider a 3-dimensional vine copula model with Gaussian pair-copulas as given in the structure matrix $M_3$ and the family matrix $\mathcal{B}^{Gauss}$. We use the parameter matrix $\btheta^{Gauss}$ to denote the dependence parameters of the bivariate Gaussian copulas.
\[
M_3=\begin{pmatrix}
3 \\
1 & 2 \\
2 & 1 & 1 \\
\end{pmatrix}\
\mathcal{B}^{Gauss}=\begin{pmatrix}
& \\
Gauss & \\
Gauss & Gauss & \ \\
\end{pmatrix}\
\btheta^{Gauss}=\begin{pmatrix}
& \\
0.34 & \\
0.79 & 0.35 & \ \\
\end{pmatrix}
\]
Then, we can calculate the asymptotic standard errors for each parameter based on the expected information matrix $\cali(\btheta^{Gauss})$ or rather $\boldsymbol{V}^{dep}$ for the MLE case (see Equation (\ref{fisher_information})) and for the sequential estimation case (Equation (\ref{seqerrors}) and \citet{haff2010}), respectively. The order of the entries in the according asymptotic standard error matrices $\boldsymbol{ASE}^{MLE}$ and $\boldsymbol{ASE}^{seq}$ are the same as in the parameter matrix $\boldsymbol{\theta}^{Gauss}$. For $\cali(\btheta^{Gauss})$ and $\boldsymbol{V}^{dep}$, the order of the parameters is $(\rho_{12},\rho_{23},\rho_{13|2})$.
{\footnotesize
\[
\cali^{MLE}(\btheta^{Gauss}) = 
\mathbbm{E}_{\boldsymbol \theta}\Bigg[-\Big( \frac{\partial^2}{\partial \theta_i^{Gauss} \partial \theta_j^{Gauss} } l(\btheta^{Gauss}) \Big)_{i,j=1,\ldots,3}\Bigg] = 
\begin{pmatrix}[rrr]
1.62 &  -0.77 & 0.15 \\
-0.77 & 12.40 & 0.80 \\
0.15 & 0.80 & 1.42 \\
\end{pmatrix}
\]
}
\[
\boldsymbol{V}^{dep,MLE} = \left(\cali^{MLE}\right)^{-1}(\btheta^{Gauss}) =
\begin{pmatrix}
\hphantom{-}0.65 & \hphantom{-}0.05 & -0.10 \\
\hphantom{-}0.05 & \hphantom{-}0.09 & -0.05 \\
-0.10 & -0.05 & \hphantom{-}0.75 \\
\end{pmatrix}.
\]
This implies that the asymptotic standard errors are given by the square roots of the diagonal elements as
\[
\boldsymbol{ASE}^{MLE} =\begin{pmatrix}
& \\
0.86 & \\
0.29 & 0.80 & \ \\
\end{pmatrix}, \ \ \
\boldsymbol{ASE}^{seq}=\begin{pmatrix}
& \\
0.89 & \\
0.31 & 0.83 & \ \\
\end{pmatrix}.
\]

For the Gaussian distribution the occurring integrals can be computed analytically, too, see \ref{appendix:Gauss}. The results indicate that regardless of the applied estimation method approximately 100 observations are required to estimate the parameters of the 3-dimensional Gaussian copula up to $\sigma=0.01$.\\
In the second setting we change the bivariate copula families to Student t-copulas. The corresponding copula parameters are stored in $\btheta^{Student}$, where the lower triangle gives the correlation parameter (parameter 1) and the upper triangle gives the degrees of freedom (parameter 2). As before $\boldsymbol{ASE}^{MLE}$ and $\boldsymbol{ASE}^{seq}$ denote the corresponding standard errors based on Equation (\ref{eq:asymptotic_MLE}) for the full ML estimation and Equation (\ref{seqerrors}) for sequential estimation, respectively.
\begin{equation}
\label{3dimt}
\btheta^{Student}=\begin{pmatrix}
 & 3 & 3 \\
0.34 &  & 3 \\
0.79 & 0.35 &  \\
\end{pmatrix}
\end{equation}
\begin{equation*}
\boldsymbol{ASE}^{MLE}=\begin{pmatrix}
 & 12 & 12 \\
1.04 &  & 11\\
0.39 & 0.97 &  \\
\end{pmatrix} \
\boldsymbol{ASE}^{seq}=\begin{pmatrix}
 & 12 & 14 \\
1.04 &  & 12 \\
0.48 & 1.15 &  \\
\end{pmatrix}
\end{equation*}\noindent
Note that our results show that for uniform $[0,1]$ marginal distributions, the sequential procedure is less efficient than full MLE. This is interesting in combination with Theorem 2 in \citet{haff2010} which states that together with non-parametric estimation of the marginal distributions, the sequential estimation procedure for the Gaussian distribution is asymptotically as efficient as the full MLE.
\end{example}

\section{Application: Rolling window analysis of exchange rate data}
\label{sec:application}

In this section, we apply the methods detailed above to the exchange rate data analyzed by \citet{CzadoSchepsmeierMin2011} and \citet{stoeber2011c}. 

The data consists of 8 daily exchange rates quoted with respect to the US dollar during the period from July 22, 2005 to July 17, 2009, resulting in 1007 data points in total. For simplicity, we use the following abbreviations: 1=AUD (Australian dollar), 2=JPY
(Japanese yen), 3=BRL (Brazilian real), 4=CAD (Canadian dollar), 5=EUR (Euro), 6=CHF (Swiss frank), 7=INR (Indian rupee) and 8=GBP (British pound).\\
As marginal models we choose the time series models described by \citet[Chapter 5]{Schepsmeier}, which are of ARMA(P,Q)-GARCH(p,q) type. To obtain marginally uniformly distributed copula data on $[0,1]^8$, the resulting standardized residuals are transformed using the non-parametric rank transformation (see \citet{genest1995}). We could also employ the probability integral transformation based on the parametric error distributions (IFM, \citet{joe1996}) but since we are only interested in dependence properties here, we choose the non-parametric alternative which is more robust with respect to misspecification of marginal error distributions.
We perform a two-step analysis, i.e. from now on we assume the marginal distributions to be known beforehand and only consider dependence analysis based on the obtained copula data. Given the marginals, the sequential model selection procedure detailed in \citet{DissmannBrechmannCzadoKurowicka2011} suggests the R-vine described in the model matrix M (Equation (\ref{eq:matrixM})) and pictured in Figure \ref{fig:pre-rvine} as an adequate model for the whole dataset. More details on the involved bivariate copula families can be found in \ref{appendix:densities}.
The $8$-dimensional R-vine model contains 28 pair copula densities; 8 of the conditional bivariate margins associated with the vine can be modeled by independence copulas and for 7 of the remaining margins two-parametric Student-t copulas are chosen (see $\mathcal{B}$, Equation (\ref{eq:Bmatrix}), with corresponding copula parameter estimates $\hat{\boldsymbol \theta}^{MLE}$, Equation (\ref{eq:ThetaMatrix}) )
Thus, the copula model has 27 parameters in total and we obtain an observed information matrix $\cali_n(\hat{\boldsymbol \theta}^{MLE})$ of dimension $27\times 27$, where the parameters are ordered by column from lower right to upper left, i.e. $\boldsymbol{\theta}^{MLE}=(0.3, 0.48, 0.63, 0.54, \ldots)$. From this matrix, we can again obtain standard errors $\boldsymbol{SE}_n^{MLE}$ (Equation (\ref{eq:SigmaMatrix})) for our parameter estimates by computing the square roots of the diagonal elements. As in Equation (\ref{3dimt}), the estimates and corresponding standard errors for the degrees of freedom parameters of Student-t copulas can be found in upper triangular part of matrix (\ref{eq:ThetaMatrix}) and (\ref{eq:SigmaMatrix}).

{\footnotesize\begin{equation}
\hat{\boldsymbol\theta}^{MLE} = 
\begin{pmatrix}[rrrrrrrr]
 &  &  &  &  &  & 11.85 & 8.96 \\
 &  &  &  &  &  &  & 7.74 \\
 &  &  &  &  &  &  & 3.76 \\
 & -0.70 & -0.09 &  &  &  &  & 9.97 \\
 & -0.88 & -1.44 &  &  &  & & 8.41 \\
0.07 & -1.10 & -0.73 &  &  &  &  & 7.46 \\
0.26 & -1.23 & -1.17 & 1.13 & 1.08 & 0.63 &  &  \\
0.72 & 0.55 & 0.88 & 0.63 & 0.54 & 0.48 &  0.3 &  \\
\end{pmatrix}
\label{eq:ThetaMatrix}
\end{equation}}

{\footnotesize\begin{equation}
\mathcal{B} = 
\begin{pmatrix}
&  \\
 \text{Indep.}&  \\
\text{Indep.} & \text{Indep.}\\
\text{Indep.} & \text{Frank} & \text{Gauss} \\
 \text{Indep.} & \text{Frank} & \text{Frank} & \text{Indep.}  \\
\text{Gauss} & \text{r.\ Joe} & \text{Frank} & \text{Indep.} & \text{Indep.} \\
\text{Student-t} & \text{r.\ Gumbel} & \text{r.\ Gumbel} & \text{Gumbel} & \text{Frank} & \text{Frank} \\
\text{Student-t} & \text{Student-t} & \text{Student-t} & \text{Student-t} & \text{Student-t} & \text{Student-t} & \text{Gauss} \\
\end{pmatrix}
\label{eq:Bmatrix}
\end{equation}}

{\footnotesize \begin{equation}
	\cali_n(\hat{\boldsymbol \theta}^{MLE}) =
	\begin{pmatrix}[rrrrrrrr]
 1333.58 &  -88.15 &  1.39 &    8.54 & -0.07 &   -6.54 &   -0.25 & \ldots \\
  -88.15 & 1843.69 &  5.46 & -185.51 &  8.41 &  -27.61 &   -1.55 & \ldots \\
   +1.39 &    5.46 & 27.09 &   -0.19 &  0.04 &   -0.28 &    0.08 & \ldots \\
   +8.54 & -185.51 & -0.19 & 2279.43 &  0.74 & -359.39 &  200.48 & \ldots \\
   -0.07 &    8.41 &  0.04 &    0.74 & 29.21 &    2.80 &    0.75 & \ldots \\
   -6.54 &  -27.61 & -0.28 & -359.39 &  2.80 & 3824.82 &  374.15 & \ldots \\
   -0.25 &   -1.55 &  0.08 &  200.48 &  0.75 &  374.15 & 1803.30 & \ldots \\
    \vdots & \vdots & \vdots & \vdots & \vdots & \vdots & \vdots & \ddots \\
	\end{pmatrix}
\end{equation}}
{\footnotesize\begin{equation}
{\boldsymbol{SE}}_n^{MLE} = 
\begin{pmatrix}
 &  &  &  &  &  & 5.00 & 2.54 \\
 &  &  &  &  &  &  & 1.95 \\
 &  &  &  &  &  &  & 0.59 \\
 & 0.19 & 0.03 & &  & &  & 3.04 \\
 & 0.19 & 0.20 &  &  &  &  & 2.40 \\
0.03 & 0.03 & 0.19 &  &  &  &  & 2.01 \\
0.03 & 0.03 & 0.03 & 0.02 & 0.19 & 0.19 &  &  \\
 0.01 & 0.02 & 0.01 & 0.02 & 0.02 & 0.02 & 0.03 &  \\
\end{pmatrix}.
\label{eq:SigmaMatrix}
\end{equation}}

\begin{figure}[htb]
	\centering
		\includegraphics[width=0.78\textwidth]{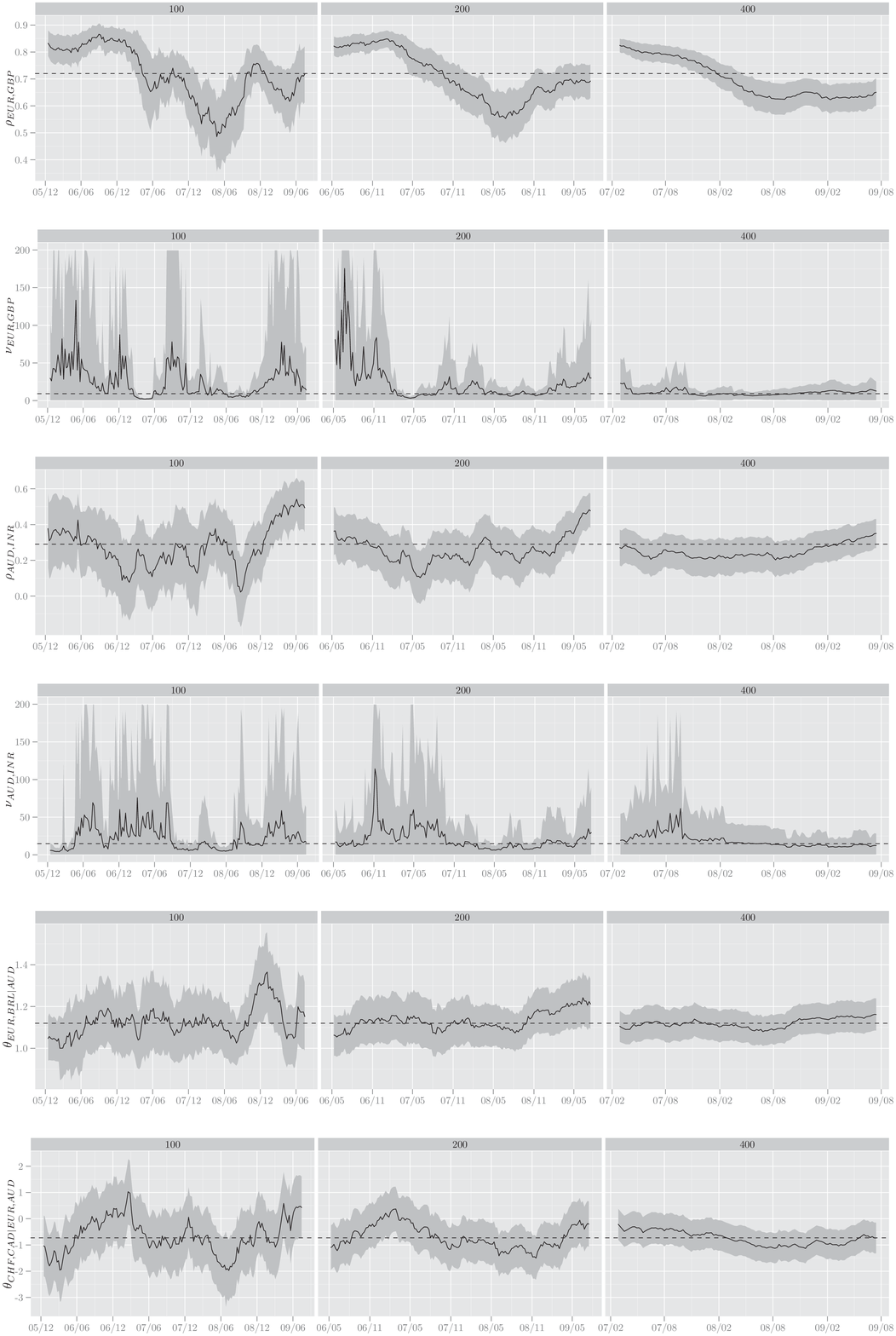}
		\vspace{-0.3cm}
		\captionsetup{justification=raggedright, font=small}
	\caption{Rolling window analysis for the exchange rate data with window size 100 (left), 200 (middle) and 400 (right) for some selected par-copulas with different copula families (t-copula (row 1-4), Gumbel (row 5) and Frank (row 6)). The x-axis indicates the endpoint of each window, with the corresponding parameter estimate on the y-axis. The dashed horizontal line in each plot is the MLE corresponding to the whole dataset.}
	\label{fig:rollingWindow}
\end{figure}

We see that the parameters of selected Gaussian copulas are significantly non-zero. For all pair-copulas which are not the independence copula there is significant dependence and also the estimated degree of freedom parameters show that the corresponding conditional copulas are significantly non-Gaussian.
To study possible parameter inhomogeneities in time, we apply a rolling window analysis as follows:
A window with a window-size of 100, 200 and 400 data points, respectively, is run over the data with step-size five, i.e. the window is moved by five trading days in each step.
For each window dataset under investigation we estimate the R-vine parameters using ML while keeping the R-vine structure $\mathcal{V}$ given by (\ref{eq:matrixM}) and the copula families $\mathcal{B}$ given by (\ref{eq:Bmatrix}) fixed. Additionally, we compute the observed information for each window and use it to obtain standard errors for the parameter estimates. 
In some windows the estimated degrees of freedom parameters are very high leading to numerical instabilities and indicating that the Student-t copulas associated with some of the bivariate margins might not be appropriate for the whole dataset.
In Figure \ref{fig:rollingWindow} we illustrate the estimated copula parameters together with pointwise approximate confidence intervals given by $[\hat{\theta}-2\hat{\sigma},\hat{\theta}+2\hat{\sigma}]$ for some selected pair-copulas.

Our parameter estimates illustrate that the dependence structure of exchange rates in fact varied over the observation period although not as much as one might guess from a naive rolling window analysis without taking into account the estimation uncertainty. We see that the amplitude of observed variation depends on the window size with smaller windows leading to more prominent peaks. However, the greater uncertainty in parameter estimates due to less observations per sample leads to bigger confidence bands which will usually cover the parameter estimates obtained from bigger windows. Given our setting with a 8-dimensional financial dataset, 100-200 observations are insufficient to estimate copula parameters with satisfactory accuracy to detect short-term fluctuations in dependence parameters (see Figure \ref{fig:rollingWindow}, row 5, with a peak in dependence visible in the results for 100 observations although we have no indication that the parameter varied over time considering standard errors and the results for larger rolling windows). Comparing results obtained with different window sizes, however, there is evidence that some dependence parameters vary over time. In the dataset at hand, the observable changes in parameter values mostly occurred during the years 2007 and 2008 in which the breakout of the subprime crisis lead to severe interruptions in financial markets and changed paradigms in international cash flows. 
The increase of volatility is reflected in weaker dependence between currency pairs, e.g. the dependence between $GBP/USD$ and $EUR/USD$ decreased to 75\% of its initial value in the year 2005 (see Figure \ref{fig:rollingWindow}, row 1). This is similar to the observations for stock markets, where dependencies during times of economic downturn are weaker than during times of economic upturn (\citet{stoeber2011c}).

\section{Discussion}
\label{sec:discussion}

While ML inference for R-vine copula based models is broadly discussed in
the literature and applied to data from various scientific disciplines, no
algorithms for the computation of the score function or observed
information have been available.

The methodological contribution of this paper is to close this gap and to allow
researchers to compute standard errors, which are at the core of ML
analysis, in a routine manner for R-vine copulas. 
Furthermore, combining our newly developed algorithms with numerical
integration techniques allows to compute the Fisher information matrix and
other quantities related to the asymptotic theory of ML estimation techniques
and thus to compare the efficiency of different estimation methods
quantitatively for a given copula model.

In a rolling window analysis we illustrate that the headline question of this paper is difficult to answer given the typical number of observations for datasets under investigation. Comparing estimation results for different rolling window sizes and taking into account pointwise confidence intervals, we obtain clear indications that the strength of dependence between some currency pairs (e.g. $GBP/USD$ - $EUR/USD$) varied over the observation period while it remained constant for others (e.g. $EUR/USD$ and $BRL/USD$ given $AUD/USD$). While the results for a window size of 100 trading days suggest that further short-term fluctuations in dependence might be present in exchange rate data, these are non-significant in our modeling framework. For the detection and study of such fluctuations, an analysis based on intraday data may be helpful and should be considered in future research.
\section{Acknowledgment}

We acknowledge substantial contributions by our working group at Technische Universit\"at M\"unchen.
Numerical calculations were performed on a Linux cluster supported by DFG grant INST 95/919-1 FUGG.
The second author gratefully acknowledges the support of the TUM Graduate School's International School of Applied Mathematics,
the first author is supported by TUM's TopMath program and a research stipend provided by Allianz Deutschland AG.

\clearpage

\bibliographystyle{model2-names}
\bibliography{references}

\appendix
\section{Algorithm for the calculation of second derivatives}
\label{appendix:algorithms}

In Section \ref{sec:secondDerivative} we introduced the seven possible cases of dependence which can occur during the calculation of the second log-likelihood derivative. In the following, we illustrate these cases in detail.
In case 1 we determine
{\footnotesize\begin{equation}\label{second_derivative_first_equation}
\begin{split}
&\frac{\partial^2}{\partial \theta \partial \gamma}  \ln\left(c_{U,V\vert \textbf{Z}}\left(F_{U \vert \textbf{Z}}(u \vert \textbf{z}, \theta), F_{V \vert \textbf{Z}}(v \vert \textbf{z}, \gamma)\right)\right) \\  
&= \frac{\partial_1 \partial_2 c_{U,V\vert \textbf{Z}}\left(F_{U \vert \textbf{Z}}(u \vert \textbf{z}, \theta), F_{V \vert \textbf{Z}}(v \vert \textbf{z}, \gamma)\right)}{ c_{U,V\vert \textbf{Z}}\left(F_{U \vert \textbf{Z}}(u \vert \textbf{z}, \theta), F_{V \vert \textbf{Z}}(v \vert \textbf{z}, \gamma)\right)} \cdot \left( \frac{\partial}{\partial \theta} F_{U \vert \textbf{Z}}(u \vert \textbf{z}, \theta) \right)  \cdot \left( \frac{\partial}{\partial \gamma} F_{V \vert \textbf{Z}}(v \vert \textbf{z}, \gamma) \right) \\
 &- \left( \frac{\partial}{\partial \theta} \ln\left(c_{U,V\vert \textbf{Z}}\left(F_{U \vert \textbf{Z}}(u \vert \textbf{z}, \theta), F_{V \vert \textbf{Z}}(v \vert \textbf{z}, \gamma)\right) \right)\right)\\  \cdot & \left( \frac{\partial}{\partial \gamma} \ln\left(c_{U,V\vert \textbf{Z}}\left(F_{U \vert \textbf{Z}}(u \vert \textbf{z}, \theta), F_{V \vert \textbf{Z}}(v \vert \textbf{z}, \gamma)\right) \right) \right),
\end{split}
\end{equation}}
for case 2
{\footnotesize\begin{equation}
\begin{split}
&\frac{\partial^2}{\partial \theta \partial \gamma}  \ln\left(c_{U,V\vert \textbf{Z}}\left(F_{U \vert \textbf{Z}}(u \vert \textbf{z}, \theta), F_{V \vert \textbf{Z}}(v \vert \textbf{z}, \theta, \gamma)\right)\right) \\  
&= \frac{\partial_1 \partial_2 c_{U,V\vert \textbf{Z}}\left(F_{U \vert \textbf{Z}}(u \vert \textbf{z}, \theta), F_{V \vert \textbf{Z}}(v \vert \textbf{z}, \theta, \gamma)\right)}{ c_{U,V\vert \textbf{Z}}\left(F_{U \vert \textbf{Z}}(u \vert \textbf{z}, \theta), F_{V \vert \textbf{Z}}(v \vert \textbf{z}, \theta, \gamma)\right)} \cdot \left( \frac{\partial}{\partial \theta} F_{U \vert \textbf{Z}}(u \vert \textbf{z}, \theta) \right)  \cdot \left( \frac{\partial}{\partial \gamma} F_{V \vert \textbf{Z}}(v \vert \textbf{z}, \theta, \gamma) \right) \\
 &- \left( \frac{\partial}{\partial \theta} \ln\left(c_{U,V\vert \textbf{Z}}\left(F_{U \vert \textbf{Z}}(u \vert \textbf{z}, \theta), F_{V \vert \textbf{Z}}(v \vert \textbf{z}, \theta, \gamma)\right) \right)\right)\\
  &\cdot  \left( \frac{\partial}{\partial \gamma} \ln(\left( c_{U,V\vert \textbf{Z}}\left(F_{U \vert \textbf{Z}}(u \vert \textbf{z}, \theta), F_{V \vert \textbf{Z}}(v \vert \textbf{z}, \theta, \gamma)\right) \right) \right) \\
 &+   \frac{\partial_2 \partial_2 c_{U,V\vert \textbf{Z}}\left(F_{U \vert \textbf{Z}}(u \vert \textbf{z}, \theta), F_{V \vert \textbf{Z}}(v \vert \textbf{z}, \theta, \gamma)\right)}{ c_{U,V\vert \textbf{Z}}\left(F_{U \vert \textbf{Z}}(u \vert \textbf{z}, \theta), F_{V \vert \textbf{Z}}(v \vert \textbf{z}, \theta, \gamma)\right)}  \cdot \left( \frac{\partial}{\partial \theta } F_{V \vert \textbf{Z}}(v \vert \textbf{z}, \theta, \gamma) \right)  \cdot \left( \frac{\partial}{\partial \gamma} F_{V \vert \textbf{Z}}(v\vert \textbf{z}, \theta, \gamma ) \right) \\
&+ \frac{ \partial_2 c_{U,V\vert \textbf{Z}}\left(F_{U \vert \textbf{Z}}(u \vert \textbf{z}, \theta), F_{V \vert \textbf{Z}}(v \vert \textbf{z}, \theta, \gamma)\right)}{c_{U,V\vert \textbf{Z}}\left(F_{U \vert \textbf{Z}}(u \vert \textbf{z}, \theta), F_{V \vert \textbf{Z}}(v \vert \textbf{z}, \theta, \gamma)\right)} \cdot \left( \frac{\partial^2}{\partial \theta \partial \gamma } F_{V \vert \textbf{Z}}(v \vert \textbf{z}, \theta, \gamma) \right),
\end{split}
\end{equation}}
and case 3 yields
{\footnotesize\begin{equation}
\begin{split}
&\frac{\partial^2}{\partial \theta \partial \gamma}  \ln(c_{U,V\vert \textbf{Z}}(F_{U \vert \textbf{Z}}(u \vert \textbf{z}, \theta, \gamma), F_{V \vert \textbf{Z}}(v \vert \textbf{z}, \theta, \gamma)))   \\ 
& = \frac{\partial_1 \partial_2 c_{U,V\vert \textbf{Z}}\left(F_{U \vert \textbf{Z}}(u \vert \textbf{z}, \theta, \gamma), F_{V \vert \textbf{Z}}(v \vert \textbf{z}, \theta, \gamma)\right)}{c_{U,V\vert \textbf{Z}}\left(F_{U \vert \textbf{Z}}(u \vert \textbf{z}, \theta, \gamma), F_{V \vert \textbf{Z}}(v \vert \textbf{z}, \theta, \gamma)\right)} \cdot \left( \frac{\partial}{\partial \theta} F_{U \vert \textbf{Z}}(u \vert \textbf{z}, \theta, \gamma) \right)  \left( \frac{\partial}{\partial \gamma} F_{V \vert \textbf{Z}}(v \vert \textbf{z}, \theta, \gamma) \right) \\
 &- \left( \frac{\partial}{\partial \theta} \ln\left(c_{U,V\vert \textbf{Z}}\left(F_{U \vert \textbf{Z}}(u \vert \textbf{z}, \theta, \gamma), F_{V \vert \textbf{Z}}(v \vert \textbf{z}, \theta, \gamma)\right) \right)\right)\\
  &\cdot \left( \frac{\partial}{\partial \gamma} \ln(\left(c_{U,V\vert \textbf{Z}}\left(F_{U \vert \textbf{Z}}(u \vert \textbf{z}, \theta, \gamma), F_{V \vert \textbf{Z}}(v \vert \textbf{z}, \theta, \gamma)\right) \right) \right) \\
 &+ \frac{\partial_1 \partial_1 c_{U,V\vert \textbf{Z}}\left(F_{U \vert \textbf{Z}}(u \vert \textbf{z}, \theta, \gamma), F_{V \vert \textbf{Z}}(v \vert \textbf{z}, \theta, \gamma)\right)}{c_{U,V\vert \textbf{Z}}\left(F_{U \vert \textbf{Z}}(u \vert \textbf{z}, \theta, \gamma), F_{V \vert \textbf{Z}}(v \vert \textbf{z}, \theta, \gamma)\right)}  \cdot \left( \frac{\partial}{\partial \theta } F_{U \vert \textbf{Z}}(u \vert \textbf{z}, \theta, \gamma) \right)  \left( \frac{\partial}{\partial \gamma} F_{U \vert \textbf{Z}}(u\vert \textbf{z}, \theta, \gamma ) \right) \\
 &+  \frac{\partial_1 \partial_2 c_{U,V\vert \textbf{Z}}\left(F_{U \vert \textbf{Z}}(u \vert \textbf{z}, \theta, \gamma), F_{V \vert \textbf{Z}}(v \vert \textbf{z}, \theta, \gamma)\right)}{c_{U,V\vert \textbf{Z}}\left(F_{U \vert \textbf{Z}}(u \vert \textbf{z}, \theta, \gamma), F_{V \vert \textbf{Z}}(v \vert \textbf{z}, \theta, \gamma)\right)}  \cdot \left( \frac{\partial}{\partial \theta } F_{V \vert \textbf{Z}}(v \vert \textbf{z}, \theta, \gamma) \right)  \left( \frac{\partial}{\partial \gamma} F_{U \vert \textbf{Z}}(u\vert \textbf{z}, \theta, \gamma ) \right) \\
 &+  \frac{\partial_1 \partial_2 c_{U,V\vert \textbf{Z}}\left(F_{U \vert \textbf{Z}}(u \vert \textbf{z}, \theta, \gamma), F_{V \vert \textbf{Z}}(v \vert \textbf{z}, \theta, \gamma)\right)}{c_{U,V\vert \textbf{Z}}\left(F_{U \vert \textbf{Z}}(u \vert \textbf{z}, \theta, \gamma), F_{V \vert \textbf{Z}}(v \vert \textbf{z}, \theta, \gamma)\right)}  \cdot \left( \frac{\partial}{\partial \theta } F_{U \vert \textbf{Z}}(u \vert \textbf{z}, \theta, \gamma) \right)  \left( \frac{\partial}{\partial \gamma} F_{V \vert \textbf{Z}}(v\vert \textbf{z}, \theta, \gamma ) \right) \\
 &+  \frac{\partial_2 \partial_2 c_{U,V\vert \textbf{Z}}\left(F_{U \vert \textbf{Z}}(u \vert \textbf{z}, \theta, \gamma), F_{V \vert \textbf{Z}}(v \vert \textbf{z}, \theta, \gamma)\right)}{c_{U,V\vert \textbf{Z}}\left(F_{U \vert \textbf{Z}}(u \vert \textbf{z}, \theta, \gamma), F_{V \vert \textbf{Z}}(v \vert \textbf{z}, \theta, \gamma)\right)}  \cdot \left( \frac{\partial}{\partial \theta } F_{V \vert \textbf{Z}}(v \vert \textbf{z}, \theta, \gamma) \right)  \left( \frac{\partial}{\partial \gamma} F_{V \vert \textbf{Z}}(v\vert \textbf{z}, \theta, \gamma ) \right) \\
 &+ \frac{ \partial_1 c_{U,V\vert \textbf{Z}}\left(F_{U \vert \textbf{Z}}(u \vert \textbf{z}, \theta, \gamma), F_{V \vert \textbf{Z}}(v \vert \textbf{z}, \theta, \gamma)\right)}{c_{U,V\vert \textbf{Z}}\left(F_{U \vert \textbf{Z}}(u \vert \textbf{z}, \theta, \gamma), F_{V \vert \textbf{Z}}(v \vert \textbf{z}, \theta, \gamma)\right)} \cdot \left( \frac{\partial^2}{\partial \theta \partial \gamma } F_{V \vert \textbf{Z}}(v \vert \textbf{z}, \theta, \gamma) \right) \\
 &+ \frac{ \partial_2 c_{U,V\vert \textbf{Z}}\left(F_{U \vert \textbf{Z}}(u \vert \textbf{z}, \theta, \gamma), F_{V \vert \textbf{Z}}(v \vert \textbf{z}, \theta, \gamma)\right)}{c_{U,V\vert \textbf{Z}}\left(F_{U \vert \textbf{Z}}(u \vert \textbf{z}, \theta, \gamma), F_{V \vert \textbf{Z}}(v \vert \textbf{z}, \theta, \gamma)\right)} \cdot \left( \frac{\partial^2}{\partial \theta \partial \gamma } F_{V \vert \textbf{Z}}(v \vert \textbf{z}, \theta, \gamma) \right).
\end{split}
\end{equation}}
Similarly, we have for case 4 that
{\footnotesize\begin{equation}
\begin{split}
&\frac{\partial^2}{\partial \theta \partial \gamma}  \ln\left( c_{U,V\vert \textbf{Z}}\left(F_{U \vert \textbf{Z}}(u \vert \textbf{z}, \theta), F_{V \vert \textbf{Z}}(v \vert \textbf{z})\vert \gamma \right)\right)\\ 
&= \left( \frac{\partial}{\partial \theta} \ln\left( c_{U,V\vert \textbf{Z}}\left(F_{U \vert \textbf{Z}}(u \vert \textbf{z}, \theta), F_{V \vert \textbf{Z}}(v \vert \textbf{z})\vert \gamma \right)\right)  \right) \cdot \frac{- \partial_\gamma  c_{U,V\vert \textbf{Z}}\left(F_{U \vert \textbf{Z}}(u \vert \textbf{z}, \theta), F_{V \vert \textbf{Z}}(v \vert \textbf{z})\vert \gamma \right)}{ c_{U,V\vert \textbf{Z}}\left(F_{U \vert \textbf{Z}}(u \vert \textbf{z}, \theta), F_{V \vert \textbf{Z}}(v \vert \textbf{z})\vert \gamma \right)} \\
 &+ \frac{\partial_\gamma \partial_1  c_{U,V\vert \textbf{Z}}\left(F_{U \vert \textbf{Z}}(u \vert \textbf{z}, \theta), F_{V \vert \textbf{Z}}(v \vert \textbf{z})\vert \gamma \right)}{ c_{U,V\vert \textbf{Z}}\left(F_{U \vert \textbf{Z}}(u \vert \textbf{z}, \theta), F_{V \vert \textbf{Z}}(v \vert \textbf{z})\vert \gamma \right)} \cdot \left(  \frac{\partial}{\partial \theta} F_{U \vert \textbf{Z}}(u\vert \textbf{z}, \theta) \right),
\end{split}
\end{equation}}
and
{\footnotesize\begin{equation}
\begin{split}
&\frac{\partial^2}{\partial \theta \partial \gamma} \ln\left( c_{U,V\vert \textbf{Z}}\left(F_{U \vert \textbf{Z}}(u \vert \textbf{z}, \theta), F_{V \vert \textbf{Z}}(v \vert \textbf{z}, \theta)\vert \gamma \right)\right)\\ 
 &=  \left( \frac{\partial}{\partial \theta} \ln\left( c_{U,V\vert \textbf{Z}}\left(F_{U \vert \textbf{Z}}(u \vert \textbf{z}, \theta), F_{V \vert \textbf{Z}}(v\vert \textbf{z}, \theta )\vert \gamma \right)\right)  \right) \\
 &\cdot \frac{- \partial_\gamma  c_{U,V\vert \textbf{Z}}\left(F_{U \vert \textbf{Z}}(u \vert \textbf{z}, \theta), F_{V \vert \textbf{Z}}(v\vert \textbf{z}, \theta)\vert \gamma \right)}{ c_{U,V\vert \textbf{Z}}\left(F_{U \vert \textbf{Z}}(u \vert \textbf{z}, \theta), F_{V \vert \textbf{Z}}(v\vert \textbf{z}, \theta)\vert \gamma \right)} \\
 &+ \frac{\partial_\gamma \partial_1  c_{U,V\vert \textbf{Z}}\left(F_{U \vert \textbf{Z}}(u \vert \textbf{z}, \theta), F_{V \vert \textbf{Z}}(v\vert \textbf{z}, \theta)\vert \gamma \right)}{ c_{U,V\vert \textbf{Z}}\left(F_{U \vert \textbf{Z}}(u \vert \textbf{z}, \theta), F_{V \vert \textbf{Z}}(v\vert \textbf{z}, \theta)\vert \gamma \right)} \cdot \left(  \frac{\partial}{\partial \theta} F_{U \vert \textbf{Z}}(u\vert \textbf{z}, \theta) \right) \\
 &+  \frac{\partial_\gamma \partial_2  c_{U,V\vert \textbf{Z}}\left(F_{U \vert \textbf{Z}}(u \vert \textbf{z}, \theta), F_{V \vert \textbf{Z}}(v\vert \textbf{z}, \theta)\vert \gamma \right)}{ c_{U,V\vert \textbf{Z}}\left(F_{U \vert \textbf{Z}}(u \vert \textbf{z}, \theta), F_{V \vert \textbf{Z}}(v\vert \textbf{z}, \theta)\vert \gamma \right)} \cdot \left(  \frac{\partial}{\partial \theta} F_{V \vert \textbf{Z}}(v\vert \textbf{z}, \theta) \right) ,
\end{split}
\end{equation}}
for the fifth case.
Finally, 
{\footnotesize\begin{equation}\label{second_derivative_last_equation}
\begin{split}
&\frac{\partial^2}{\partial \theta \partial \gamma} \ln\left( c_{U,V\vert \textbf{Z}}\left(F_{U \vert \textbf{Z}}(u \vert \textbf{z}, \theta,\gamma), F_{V \vert \textbf{Z}}(v \vert \textbf{z}) \right)\right)\\ 
&=  \frac{\partial_1 \partial_1  c_{U,V\vert \textbf{Z}}\left(F_{U \vert \textbf{Z}}(u \vert \textbf{z}, \theta,\gamma), F_{V \vert \textbf{Z}}(v \vert \textbf{z}) \right) }{ c_{U,V\vert \textbf{Z}}\left(F_{U \vert \textbf{Z}}(u \vert \textbf{z}, \theta,\gamma), F_{V \vert \textbf{Z}}(v \vert \textbf{z}) \right)} \cdot \left(  \frac{\partial}{\partial \theta} F_{U \vert \textbf{Z}}(u\vert \textbf{z}, \theta,\gamma) \right) \cdot \left(  \frac{\partial}{\partial \gamma} F_{U \vert \textbf{Z}}(u\vert \textbf{z}, \theta,\gamma) \right) \\
 &-  \left(  \frac{\partial}{\partial \gamma} \ln\left( c_{U,V\vert \textbf{Z}}\left(F_{U \vert \textbf{Z}}(u \vert \textbf{z}, \theta,\gamma), F_{V \vert \textbf{Z}}(v \vert \textbf{z}) \right)\right) \right) \\
 &\cdot  \left(  \frac{\partial}{\partial \theta} \ln\left( c_{U,V\vert \textbf{Z}}\left(F_{U \vert \textbf{Z}}(u \vert \textbf{z}, \theta,\gamma), F_{V \vert \textbf{Z}}(v \vert \textbf{z}) \right)\right) \right) \\
 &+  \frac{\partial_1  c_{U,V\vert \textbf{Z}}\left(F_{U \vert \textbf{Z}}(u \vert \textbf{z}, \theta,\gamma), F_{V \vert \textbf{Z}}(v \vert \textbf{z}) \right) }{ c_{U,V\vert \textbf{Z}}\left(F_{U \vert \textbf{Z}}(u \vert \textbf{z}, \theta,\gamma), F_{V \vert \textbf{Z}}(v \vert \textbf{z}) \right)} \cdot \left(  \frac{\partial^2}{\partial \gamma \partial \theta} F_{U \vert \textbf{Z}}(u\vert \textbf{z}, \theta,\gamma) \right), \\
&\frac{\partial^2}{\partial \theta \partial \gamma}  \ln\left( c_{U,V\vert \textbf{Z}}\left(F_{U \vert \textbf{Z}}(u \vert \textbf{z}), F_{V \vert \textbf{Z}}(v \vert \textbf{z})  \vert \theta,\gamma \right)\right) \\  
&= \frac{\partial_\theta \partial_\gamma  c_{U,V\vert \textbf{Z}}\left(F_{U \vert \textbf{Z}}(u ), F_{V \vert \textbf{Z}}(v \vert \textbf{z}) \vert \theta,\gamma \right) }{ c_{U,V\vert \textbf{Z}}\left(F_{U \vert \textbf{Z}}(u \vert \textbf{z}), F_{V \vert \textbf{Z}}(v \vert \textbf{z})\vert \theta,\gamma \right)} \\  
 &- \frac{\partial_\theta  c_{U,V\vert \textbf{Z}}\left(F_{U \vert \textbf{Z}}(u ), F_{V \vert \textbf{Z}}(v \vert \textbf{z}) \vert \theta,\gamma \right) \cdot \partial_\gamma  c_{U,V\vert \textbf{Z}}\left(F_{U \vert \textbf{Z}}(u \vert \textbf{z} ), F_{V \vert \textbf{Z}}(v) \vert \theta,\gamma \right)}{ c_{U,V\vert \textbf{Z}}\left(F_{U \vert \textbf{Z}}(u \vert \textbf{z}), F_{V \vert \textbf{Z}}(v \vert \textbf{z})\vert \theta,\gamma \right)^2}. 
\end{split}
\end{equation}}

\begin{center}
  \captionsetup{style=ruled,type=algorithm,skip=0pt}
  \makeatletter
    \fst@algorithm\@fs@pre
  \makeatother
  \caption{Second derivative with respect to the parameters $\theta^{\tilde k, \tilde i}$ and $\theta^{\hat k, \hat i}$. }
  \makeatletter
    \@fs@mid
  \makeatother

\label{alg_3}
\begin{flushleft}
The input of the algorithm is a $d$-dimensional R-vine matrix $M$ with maximum matrix $\tilde M$ and parameter matrix $\btheta$, and matrices $C^{\tilde k,\tilde i}$, $C^{\hat k,\hat i}$ determined using Algorithm \ref{alg_1} for parameters $\theta^{\tilde k, \tilde i}$ and $\theta^{\hat k, \hat i}$ of the R-vine parameter matrix. Further, we assume the matrices $V^{direct}$, $V^{indirect}$ and $V^{values}$, the matrices $S1^{direct, \tilde k, \tilde i}$, $S1^{indirect, \tilde k, \tilde i}$ and $S1^{values, \tilde k, \tilde i}$ and $S1^{direct, \hat k, \hat i}$, $S1^{indirect, \hat k, \hat i}$ and $S1^{values, \hat k, \hat i}$ to be given. The output will be the value of the second derivative of the copula log-likelihood for the given observation with respect to parameters $\theta_{\tilde k, \tilde i}$ and $\theta_{\hat k, \hat i}$. Without loss of generality, we assume that $\hat i \geq \tilde i$, and $\hat k \geq \tilde k$ if $\hat i = \tilde i$.
\end{flushleft}
\begin{algorithmic}[1]
\IF{$c_{\tilde k,\tilde i}^{\hat k, \hat i} == 1$}
	\STATE Set $m= \tilde{ m}_{\tilde k, \tilde i}$
	\STATE Set $z_1= v^{direct}_{\tilde k, \tilde i},  \tilde{z}_1= s1^{direct,\hat k, \hat i}_{\tilde k, \tilde i}$
	\IF{$m==m_{\tilde k, \tilde i}$}
		\STATE Set $z_2=v^{direct}_{\tilde k, d - m + 1},  \tilde{z}_2=s1^{direct,\hat k,\hat i}_{\tilde k, d - m + 1}$
	\ELSE
		\STATE Set $z_2=v^{indirect}_{\tilde k, d - m + 1}, \tilde{z}_2=s1^{indirect, \hat k, \hat i}_{\tilde k, d - m + 1}$
	\ENDIF
	\STATE Set $s2^{direct}_{\tilde k -1,\tilde i} = 0,  s2^{indirect}_{\tilde k -1, \tilde i} = 0,  s2^{values}_{\tilde k,\tilde i} = 0$
	\IF{$\tilde k ==\hat k \ \& \ \tilde i == \hat i$}
		\STATE Set $s2^{direct}_{\tilde k -1,\tilde i} =   \partial_{\theta_{\tilde k,\tilde i}} \partial_{\theta_{\tilde k,\tilde i}} h(z_1,z_2 \vert \mathcal{B}^{\tilde k,\tilde i}, \theta^{\tilde k,\tilde i})$
		\STATE Set $s2^{indirect}_{\tilde k -1,\tilde i} =   \partial_{\theta_{\tilde k,\tilde i}} \partial_{\theta_{\tilde k,\tilde i}} h(z_2,z_1 \vert \mathcal{B}^{\tilde k,\tilde i}, \theta^{\tilde k,\tilde i})$
		\STATE Set $s2^{values}_{\tilde k,\tilde i} = \frac{\partial_{\theta_{\tilde k,\tilde i}} \partial_{\theta_{\tilde k,\tilde i}} c(z_1,z_2 \vert \mathcal{B}^{\tilde k,\tilde i}, \theta^{\tilde k,\tilde i})}{exp(v^{values}_{\tilde k, \tilde i})} - (s1^{values,\hat k,\hat i}_{\tilde k, \tilde i})^2$
	\ENDIF
	\IF{$c^{\hat k, \hat i}_{\tilde k+1,\tilde i}==1$}
		\STATE Set $s2^{values}_{\tilde k, \tilde i} = s1^{values,\hat k,\hat i}_{\tilde k, \tilde i} \cdot \frac{- \partial_{\theta_{\tilde k, \tilde i}}c(z_1,z_2 \vert \mathcal{B}^{\tilde k,\tilde i}, \theta^{\tilde k, \tilde i})}{exp(v^{values}_{\tilde k, \tilde i})} +  \frac{ \partial_1 \partial_{\theta_{\tilde k, \tilde i}}c(z_1,z_2\vert \mathcal{B}^{\tilde k,\tilde i}, \theta^{\tilde k, \tilde i})}{exp(v^{values}_{\tilde k, \tilde i})} \cdot \tilde{z_1}$
		\STATE Set $s2^{direct}_{\tilde k -1, \tilde i} = \partial_ 1 \partial_{\theta_{\tilde k,\tilde i}} h(z_1,z_2 \vert \mathcal{B}^{\tilde k,\tilde i}, \theta^{\tilde k,\tilde i}) \cdot \tilde{z_1}$
		\STATE Set $s2^{indirect}_{\tilde k -1, \tilde i} = \partial_2 \partial_{\theta_{\tilde k,\tilde i}} h(z_2,z_1 \vert \mathcal{B}^{\tilde k,\tilde i}, \theta^{\tilde k,\tilde i}) \cdot \tilde{z_1}$
	\ENDIF
	\IF{$c^{\hat k, \hat i}_{\tilde k+1,d-m+1}==1$}
		\STATE Set $s2^{values}_{\tilde k, \tilde i} =  s2^{values}_{\tilde k, \tilde i} +  \frac{ \partial_2 \partial_{\theta_{\tilde k, \tilde i}}c(z_1,z_2\vert \mathcal{B}^{\tilde k, \tilde i}, \theta^{\tilde k, \tilde i})}{exp(v^{values}_{\tilde k, \tilde i})} \cdot \tilde{z_2}$
		\IF{$c^{\hat k, \hat i}_{k+1,i}==0$}
			\STATE Set $s2^{values}_{\tilde k, \tilde i} = s2^{values}_{\tilde k, \tilde i} + s1^{values,\hat k,\hat i}_{\tilde k, \tilde i} \cdot \frac{- \partial_{\theta_{\tilde k, \tilde i}}c(z_1,z_2 \vert  \mathcal{B}^{\tilde k, \tilde i}, \theta^{\tilde k, \tilde i})}{exp(v^{values}_{\tilde k, \tilde i})}$
		\ENDIF
		\STATE Set $s2^{direct}_{\tilde k -1,\tilde i} =  s2^{direct}_{\tilde k -1,\tilde i} + \partial_ 2 \partial_{\theta_{\tilde k,\tilde i}} h(z_1,z_2 \vert  \mathcal{B}^{\tilde k,\tilde i}, \theta^{\tilde k,\tilde i}) \cdot \tilde{z_2}$
		\STATE Set $s2^{indirect}_{\tilde k -1,\tilde i} = s2^{indirect}_{\tilde k -1,\tilde i}+ \partial_1 \partial_{\theta_{\tilde k,\tilde i}} h(z_2,z_1 \vert  \mathcal{B}^{\tilde k,\tilde i}, \theta^{\tilde k,\tilde i}) \cdot \tilde{z_2}$
	\ENDIF
\ENDIF

\FOR{$i=\tilde i, \dots, 1$}
	\FOR{$k=\tilde k-1, \dots, i+1$}
		\STATE Set $m= \tilde{ m}_{k,i}$
		\STATE Set $z_1= v^{direct}_{k,i},  \tilde{z}_1^{\hat k, \hat i}= s1^{direct,\hat k, \hat i}_{k,i}, \tilde{z}_1^{\tilde k, \tilde i}= s1^{direct,\tilde k, \tilde i}_{k,  i}, \bar{z}_1=s2^{direct}_{k,i}$
		\IF{$m==m_{k, i}$}
			\STATE Set $z_2=v^{direct}_{ k, d - m + 1},  \tilde{z}_2^{\hat k, \hat i}=s1^{direct,\hat k,\hat i}_{k, d - m + 1}, \tilde{z}_2^{\tilde k, \tilde i}=s1^{direct,\tilde k,\tilde i}_{k, d - m + 1}, \bar{z}_2 = s2^{direct}_{k, d - m + 1}$
		\ELSE
			\STATE Set $z_2=v^{indirect}_{ k, d - m + 1}, \tilde{z}_2^{\hat k, \hat i}=s1^{indirect,\hat k,\hat i}_{k, d - m + 1}, \tilde{z}_2^{\tilde k, \tilde i}=s1^{indirect,\tilde k,\tilde i}_{k, d - m + 1}, \bar{z}_2 = s2^{indirect}_{k, d - m + 1}$
		\ENDIF
		\STATE Set $s2^{values}_{k,i} = - s1^{values,\hat k,\hat i}_{k, i} \cdot s1^{values,\tilde k,\tilde i}_{k, i}, s2^{direct}_{k-1,i} = 0, s2^{indirect}_{k-1,i} = 0 $
		\IF{$c^{\hat k, \hat i}_{k+1,i}==1 \ \ \& \ \ c^{\tilde k, \tilde i}_{k+1,i}==1$}
			\STATE Set $s2^{values}_{k,i} = s2^{values}_{k,i} + \frac{\partial_1 \partial_1 c(z_1,z_2 \vert \mathcal{B}^{k,i}, \theta^{k,i} )}{exp(v^{values}_{ k, i})} \cdot  \tilde{z}_1^{\hat k, \hat i} \cdot \tilde{z}_1^{\tilde k, \tilde i} + \frac{\partial_1 c(z_1,z_2 \vert \mathcal{B}^{k,i}, \theta^{k,i})}{ cop} \cdot \bar{z}_1$
			\STATE Set $s2^{direct}_{k-1,i} =  s2^{direct}_{k-1,i} + \partial_1 h(z_1, z_2 \vert \mathcal{B}^{k,i}, \theta^{k,i}) \cdot \bar{z_1} + \partial_1 \partial_1 h(z_1, z_2 \vert \mathcal{B}^{k,i}, \theta^{k,i}) \cdot \tilde{z}_1^{\hat k, \hat i} \cdot \tilde{z}_1^{\tilde k, \tilde i}$
			\STATE Set $s2^{indirect}_{k-1,i} =  s2^{indirect}_{k-1,i} + \partial_2 h(z_2, z_1 \vert \mathcal{B}^{k,i}, \theta^{k,i}) \cdot \bar{z_1} + \partial_2 \partial_2 h(z_2, z_1 \vert \mathcal{B}^{k,i}, \theta^{k,i}) \cdot \tilde{z}_1^{\hat k, \hat i} \cdot \tilde{z}_1^{\tilde k, \tilde i}$
		\ENDIF
		\IF{$c^{\hat k, \hat i}_{k+1,d-m+1}==1 \ \ \& \ \ c^{\tilde k, \tilde i}_{k+1,d-m+1}==1$}
			\STATE Set $s2^{values}_{k,i} = s2^{values}_{k,i} + \frac{\partial_2 \partial_2 c(z_1,z_2 \vert \mathcal{B}^{k,i}, \theta^{k,i} )}{exp(v^{values}_{ k, i})} \cdot  \tilde{z}_2^{\hat k, \hat i} \cdot \tilde{z}_2^{\tilde k, \tilde i} + \frac{\partial_2 c(z_1,z_2 \vert \theta_{k,i})}{ exp(v^{values}_{ k, i})} \cdot \bar{z}_2$
			\STATE Set $s2^{direct}_{k-1,i} =  s2^{direct}_{k-1,i} + \partial_2 h(z_1, z_2 \vert \mathcal{B}^{k,i}, \theta^{k,i}) \cdot \bar{z_2} + \partial_2 \partial_2 h(z_1, z_2 \vert \mathcal{B}^{k,i}, \theta^{k,i}) \cdot \tilde{z}_2^{\hat k, \hat i} \cdot \tilde{z}_2^{\tilde k, \tilde i}$
			\STATE Set $s2^{indirect}_{k-1,i} =  s2^{indirect}_{k-1,i} + \partial_1 h(z_2, z_1 \vert \mathcal{B}^{k,i}, \theta^{k,i}) \cdot \bar{z_2} + \partial_1 \partial_1 h(z_2, z_1 \vert \mathcal{B}^{k,i}, \theta^{k,i}) \cdot \tilde{z}_2^{\hat k, \hat i} \cdot \tilde{z}_2^{\tilde k, \tilde i}$
		\ENDIF
		\IF{$c^{\hat k, \hat i}_{k+1,i}==1 \ \ \& \ \ c^{\tilde k, \tilde i}_{k+1,d-m+1}==1$}
			\STATE Set $s2^{values}_{k,i} = s2^{values}_{k,i} + \frac{\partial_1 \partial_2 c(z_1,z_2 \vert \mathcal{B}^{k,i}, \theta^{k,i} )}{exp(v^{values}_{ k, i})} \cdot  \tilde{z}_1^{\hat k, \hat i} \cdot \tilde{z}_2^{\tilde k, \tilde i}$
			\STATE Set $s2^{direct}_{k-1,i} =  s2^{direct}_{k-1,i} +  \partial_1 \partial_2 h(z_1, z_2 \vert\mathcal{B}^{k,i}, \theta^{k,i}) \cdot \tilde{z}_1^{\hat k, \hat i} \cdot \tilde{z}_2^{\tilde k, \tilde i}$
			\STATE Set $s2^{indirect}_{k-1,i} =  s2^{direct}_{k-1,i} +  \partial_1 \partial_2 h(z_2, z_1 \vert \mathcal{B}^{k,i}, \theta^{k,i}) \cdot \tilde{z}_1^{\hat k, \hat i} \cdot \tilde{z}_2^{\tilde k, \tilde i}$
		\ENDIF
		\IF{$c^{\hat k, \hat i}_{k+1,d-m+1}==1 \ \ \& \ \ c^{\tilde k, \tilde i}_{k+1,i}==1$}
			\STATE Set $s2^{values}_{k,i} = s2^{values}_{k,i} + \frac{\partial_2 \partial_1 c(z_1,z_2 \vert \mathcal{B}^{k,i}, \theta^{k,i} )}{exp(v^{values}_{ k, i})} \cdot  \tilde{z}_2^{\hat k, \hat i} \cdot \tilde{z}_1^{\tilde k, \tilde i}$
			\STATE Set $s2^{direct}_{k-1,i} =  s2^{direct}_{k-1,i} +  \partial_1 \partial_2 h(z_1, z_2 \vert \mathcal{B}^{k,i}, \theta^{k,i}) \cdot \tilde{z}_2^{\hat k, \hat i} \cdot \tilde{z}_1^{\tilde k, \tilde i}$
			\STATE Set $s2^{indirect}_{k-1,i} =  s2^{direct}_{k-1,i} +  \partial_1 \partial_2 h(z_2, z_1 \vert \mathcal{B}^{k,i}, \theta^{k,i}) \cdot \tilde{z}_2^{\hat k, \hat i} \cdot \tilde{z}_1^{\tilde k, \tilde i}$
		\ENDIF
	\ENDFOR
\ENDFOR
\RETURN $\sum_{k,i = 1, \dots,d} s2^{values}_{k,i}$

\end{algorithmic}
\makeatletter
    \@fs@post
  \makeatother
\end{center}


\section{Calculation of the covariance matrix in the Gaussian case}
\label{appendix:Gauss}

While analytical results on the Fisher information for the multivariate normal distribution are well known (\citet{MardiaMarshall1984}) we will now illustrate how the matrices $\boldsymbol{\calk}_{\theta}$ and $\boldsymbol{\calj}_{\theta}$ (Equation (\ref{eq:matrixK}) and (\ref{eq:matrixJ})) can be calculated. We consider a 3-dimensional Gaussian distribution

\begin{equation*}
\begin{pmatrix}
X_1\\
X_2\\
X_3\\
\end{pmatrix} \sim N_3(\boldsymbol{0},\boldsymbol{\Sigma}), \ \ \ \boldsymbol{\Sigma} =
\begin{pmatrix}
1 & \rho_{12} & \rho_{13} \\
\rho_{12} & 1 & \rho_{23} \\
\rho_{13} & \rho_{23} & 1 \\
\end{pmatrix},
\end{equation*}

with density $f_{123}$ and corresponding copula $c_{123}$.
Exampli gratia, we show the computation for the entry $(2,1)$ in $\boldsymbol{\calk}_{\theta}$ in detail. The other entries in $\boldsymbol{\calk}_{\theta}$ and $\boldsymbol{\calj}_{\theta}$ are obtained similarly.
The first step is to calculate the following integral:

{\footnotesize\begin{equation}
	\int_{[0,1]^{3}} \left(\frac{\partial}{\partial\rho_{12}} \ln(c_{12}(u_1,u_2|\rho_{12}))\right)\left(\frac{\partial}{\partial\rho_{23}} \ln(c_{23}(u_2,u_3|\rho_{23}))\right) c_{123}(u_1,u_2,u_3) du_1du_2 du_3,
	\label{eq:integral}
\end{equation}}\noindent
where $c_{12}$ and $c_{23}$ are the corresponding copulas to the bivariate marginal distributions $f_{12}$ and $f_{23}$, respectively. Since the integral is independent of the univariate marginal distributions, we can compute it using standard normal margins (see \citet{Smith2007}):

{\footnotesize\begin{equation}
	\int_{\IR^3} \left(\frac{\partial}{\partial\rho_{12}} \ln(f_{12}(x_1,x_2|\rho_{12}))\right)\left(\frac{\partial}{\partial\rho_{23}} \ln(f_{23}(x_2,x_3|\rho_{23}))\right) f_{123}(x_1,x_2,x_3) dx_1dx_2dx_3,
	\label{eq:integral2}
\end{equation}}\noindent
where $f_{12}$ and $f_{23}$ are the according bivariate normal distributions. The 3-dimensional and bivariate normal densities in (\ref{eq:integral}) and (\ref{eq:integral2}) can be expressed as 

{\footnotesize\begin{equation}
\begin{split}
	&f_{123}(x_1,x_2,x_3) = \frac{\sqrt{2}}{{\pi}^{3/2}\sqrt{2\,\rho_{13}\rho_{12}\rho
_{23}-{\rho_{13}}^{2}-{\rho_{12}}^{2}+1-{\rho_{23}}^{2}}} \\
&\cdot\exp\left\{-\frac{1}{2}{\frac {-{x_{{1}}}^{2}-{x_{{2}}}^{2}-{x_{{3}}}^{2}
+{x_{{1}}}^{2}{\rho_{23}}^{2}
+{x_{{2}}}^{2}{\rho_{13}}^{2}
+{x_{{3}}}^{2}{\rho_{12}}^{2}
+2\,x_{{1}}x_{{2}}\rho_{12}
+2\,x_{{1}}x_{{3}}\rho_{13} }
{-2\,\rho_{13}\rho_{12}\rho_{23}+{\rho_{13}}^{2}+{\rho_{12}}^{2}-1+{\rho_{23}}^{2}}}\right\} \\
&\cdot\exp\left\{\frac{
+2\,x_{{2}}x_{{3}}\rho_{23}
-2\,x_{{1}}x_{{2}}\rho_{13}\rho_{23}
-2\,x_{{1}}x_{{3}}\rho_{12}\rho_{23}
-2\,x_{{2}}x_{{3}}\rho_{13}\rho_{12}}
{-2\,\rho_{13}\rho_{12}\rho_{23}+{\rho_{13}}^{2}+{\rho_{12}}^{2}-1+{\rho_{23}}^{2}}\right\}
\end{split}
\label{eq:3dimGauss}
\end{equation}}
and

\begin{equation}
	f_{12}(x_1,x_2) = \frac{1}{2\pi}{\frac {1}{\sqrt {1-{\rho_{12}}^{2}}}}
	\,{\exp\left\{-\frac{1}{2}\,{\frac {-{x_{{1}}}^{2}+2\,x_{{1}}x_{{2}}\rho_{12}
	-{x_{{2}}}^{2}}{ \left( -1+\rho_{12} \right)  \left( \rho_{12}+1 \right) }}\right\}}.
\end{equation}\noindent
Further, the derivatives needed in Equation (\ref{eq:integral2}) are
\begin{equation}
	\frac{\partial}{\partial \rho_{12}} \ln(f_{12}(x_1,x_2|\rho_{12})) = 
	-{\frac {
	{\rho_{12}}^{3}-x_{{1}}x_{{2}}{\rho_{12}}^{2}+{x_{{2}}}^{2}\rho_{12}
	-\rho_{12}+{x_{{1}}}^{2}\rho_{12}-x_{{1}}x_{{2}} }
	{ \left( -1+\rho_{12} \right) ^{2} \left( \rho_{12}+1 \right) ^{2}} }
	\label{eq:ableitung1}
\end{equation}
and
\begin{equation}
	\frac{\partial}{\partial \rho_{23}} \ln(f_{23}(x_2,x_3|\rho_{23})) =
	-{\frac {
	{\rho_{23}}^{3}-x_{{2}}x_{{3}}{\rho_{23}}^{2}+{x_{{3}}}^{2}\rho_{23}
	-\rho_{23}+{x_{{2}}}^{2}\rho_{23}-x_{{2}}x_{{3}} }
	{ \left( -1+\rho_{23} \right) ^{2} \left( \rho_{23}+1 \right) ^{2}} }.
	\label{eq:ableitung2}
\end{equation}
Using (\ref{eq:3dimGauss}), (\ref{eq:ableitung1}) and (\ref{eq:ableitung2}) in (\ref{eq:integral2}) we get

{\footnotesize\begin{equation}
\begin{split}
	(\ref{eq:integral2}) &= \int_{\IR^3}
	{\frac {
	{\rho_{12}}^{3}-x_{{1}}x_{{2}}{\rho_{12}}^{2}+{x_{{2}}}^{2}\rho_{12}
	-\rho_{12}+{x_{{1}}}^{2}\rho_{12}-x_{{1}}x_{{2}} }
	{ \left( -1+\rho_{12} \right) ^{2} \left( \rho_{12}+1 \right) ^{2}} } \\
	&\cdot {\frac {
	{\rho_{23}}^{3}-x_{{2}}x_{{3}}{\rho_{23}}^{2}+{x_{{3}}}^{2}\rho_{23}
	-\rho_{23}+{x_{{2}}}^{2}\rho_{23}-x_{{2}}x_{{3}} }
	{ \left( -1+\rho_{23} \right) ^{2} \left( \rho_{23}+1 \right) ^{2}} } 
\cdot f_{123}(x_1,x_2,x_3)
dx_1dx_2dx_3.
\end{split}
\label{eq:7}
\end{equation}}\noindent
The integral (\ref{eq:7}) can be solved using well known results on product moments of multivariate normal distributions (see \citet{Isserlis1918}).
\begin{equation}
\begin{split}
	(\ref{eq:7}) &= \frac {
	\rho_{23}{\rho_{12}}^{3}+{\rho_{12}}^{3}{\rho_{23}}^{3}
	-3\,{\rho_{23}}^{2}{\rho_{12}}^{2}\rho_{13}-{\rho_{12}}^{2}\rho_{13}
	+2\,\rho_{23}{\rho_{13}}^{2}\rho_{12} }
	{ \left( \rho_{23}+1 \right) ^{2} \left( -1+\rho_{23} \right) ^{2} \left( -1+{\rho_{12}}^{2} \right) ^{2} } \\
	&+\frac{-\rho_{12}\rho_{23}+{\rho_{23}}^{3}\rho_{12}+\rho_{13}-{\rho_{23}}^{2}\rho_{13} }
{ \left( \rho_{23}+1 \right) ^{2} \left( -1+\rho_{23} \right) ^{2} \left( -1+{\rho_{12}}^{2} \right) ^{2} }
\label{eq:8}
\end{split}
\end{equation}
Since $\left( \rho_{23}+1 \right) ^{2} \left( -1+\rho_{23} \right) ^{2} = (1-\rho_{23}^2)^2$ we can simplify Equation (\ref{eq:8}) to
{\footnotesize
\begin{equation*}
\begin{split}
	(\ref{eq:integral}) &= (\ref{eq:8}) = 
	\frac{(\rho_{13}-\rho_{12}\rho_{23})(1-\rho_{12}^2)(1-\rho_{23}^2) 
	+ 2\rho_{12}\rho_{23}(\rho_{13}-\rho_{12}\rho_{23})^2 }
	{ (1-\rho_{12}^2)^2(1-\rho_{23}^2)^2 } \\
	&= \frac{\rho_{13}-\rho_{12}\rho_{23}}{(1-\rho_{12}^2)(1-\rho_{23}^2)}
	+2\rho_{12}\rho_{23}\frac{(\rho_{13}-\rho_{12}\rho_{23})^2}{ (1-\rho_{12}^2)^2(1-\rho_{23}^2)^2} \\
	&= \frac{k_{12}}{(1-\rho_{12}^2)(1-\rho_{23}^2)},
\end{split}
\end{equation*}}\noindent
with
\begin{equation*}
	k_{12} = (\rho_{13}-\rho_{12}\rho_{23})\left(1+2\rho_{12}\rho_{23}\frac{\rho_{13}-\rho_{12}\rho_{23}}{(1-\rho_{12}^2)(1-\rho_{23}^2)}\right).
\end{equation*}

\noindent
For the computation of terms corresponding to parameter $\rho_{13|2}$, note that 
\[
	\rho_{13|2} = \frac{\rho_{13} - \rho_{12}\rho_{23}}{\sqrt{(1-\rho_{12}^2)(1-\rho_{23}^2)}}
\]
and
\begin{equation*}
	\rho_{13} = \rho_{13|2}\sqrt{(1-\rho_{12}^2)(1-\rho_{23}^2)} + \rho_{12}\rho_{23},
\label{eq:rho13}
\end{equation*}
which means that (\ref{eq:3dimGauss}), (\ref{eq:ableitung1}) and (\ref{eq:ableitung2}) can easily be re-parametrized.\\
The final matrices are
\[
\boldsymbol{\calk}_{\theta} = \begin{pmatrix}
\frac{1+\rho_{12}^2}{(1-\rho_{12}^2)^2} & \frac{k_{12}}{(1-\rho_{12}^2)(1-\rho_{23}^2)} & 0 \\
\frac{k_{12}}{(1-\rho_{12}^2)(1-\rho_{23}^2)} & \frac{1+\rho_{23}^2}{(1-\rho_{23}^2)^2} & 0 \\
0 & 0 & \frac{1+\rho_{13|2}}{(\rho_{13|2}^2-1)^2} \\
\end{pmatrix}
\]
\[
\boldsymbol{\calj}_{\theta} = \begin{pmatrix}
\frac{1+\rho_{12}^2}{(1-\rho_{12}^2)^2} & 0 & 0 \\
0 & \frac{1+\rho_{23}^2}{(1-\rho_{23}^2)^2} & 0 \\
\frac{\rho_{13|2}\rho_{12}}{(\rho_{12}^2-1)(\rho_{13|2}^2-1)} & \frac{\rho_{13|2}\rho_{23}}{(\rho_{23}^2-1)(\rho_{13|2}^2-1)} & \frac{1+\rho_{13|2}}{(\rho_{13|2}^2-1)^2} \\
\end{pmatrix}.
\]

\section{Bivariate copula densities}
\label{appendix:densities}

This appendix introduces the bivariate copula families which are included in our examples and applications. For more details we refer to \citet{SchepsmeierStoeber2012}.

The Gaussian copula with correlation parameter $\rho\in(-1,1)$ is defined by its density 
\begin{equation}
	c_{Gauss}(u_1,u_2|\rho) = \frac{1}{\sqrt{1-\rho^2}}\exp\left\{-\frac{\rho^2(x_1^2+x_2^2)-2\rho x_1x_2}{2(1-\rho^2)}\right\},
\label{eq:gaussDensity}
\end{equation}
\noindent
where $x_i = \Phi^{-1}(u_i)$, $i=1,2$ with $\Phi$ being the standard normal cdf and $\Phi^{-1}$ (the quantile function) its functional inverse.
Denoting the density of the univariate Student-t distribution with degrees of freedom $\nu$ as 
\[
	dt(x|\nu) = \frac{\Gamma\left(\frac{\nu+1}{2}\right)}{\Gamma\left(\frac{\nu}{2}\right)\sqrt{\pi\nu}} \left(1+\frac{x^2}{\nu}\right)^{-\frac{\nu+1}{2}},
\]
the bivariate Student-t copula's density is given by
{\footnotesize
\begin{equation*}
	c_{Student-t}(u_1,u_2|\rho,\nu) = \frac{1}{2\pi\sqrt{1-\rho^2}}\frac{1}{dt(x_1,\nu)dt(x_2,\nu)}\left(1+\frac{x_1^2+x_2^2-2\rho x_1 x_2}{\nu(1-\rho^2)}\right)^{-\frac{\nu+2}{2}}.
\label{eq:tDensity}
\end{equation*}}\noindent
Again, $x_i$ is given as $x_i:=t_{\nu}^{-1}(u_i)$, $i=1,2$, with $t_{\nu}^{-1}(\cdot)$ now being the quantile function of the univariate Student-t distribution. 

In addition to these two well-known members of the elliptical class, we use the Gumbel copula, its rotated version to cover negative dependence, the Frank copula and the rotated Joe copula, which are so-called Archimedean copulas.

The density of the Gumbel copula with parameter $\theta \geq 1$ is given as 
{\small
\begin{equation*}
\begin{split}
	c_{Gumbel}(u_1,u_2|\theta) &= C(u_1,u_2|\theta)(u_1 u_2)^{-1}\{(-\ln(u_1))^{\theta}+(-\ln(u_2))^{\theta}\}^{-2+\frac{2}{\theta}} \\
	&\cdot (\ln(u_1) \ln(u_2))^{\theta-1}\cdot  \{1+(\theta-1)((-\ln(u_1))^{\theta}+(-\ln(u_2))^{\theta})^{-\frac{1}{\theta}}\} \\
\end{split}
\end{equation*}}\noindent
with
\begin{equation*}
 C_{Gumbel}(u_1,u_2|\theta) = \exp [-\{(-\ln(u_1))^{\theta}+(-\ln(u_2))^{\theta}\}^{\frac{1}{\theta}}],
\end{equation*}
being the corresponding cumulative distribution function.
While the Gumbel copula can only cover positive dependence, it can be rotated to fit negatively dependent data:
\begin{equation*}
	c_{rotated}(u_1,u_2):=c(u_1,1-u_2).
\label{eq:rotated}
\end{equation*}
Further, the Frank copula with parameter $\theta \in [-\infty,\infty]\backslash\{0\}$ has density 
{\small
\begin{equation*}
c_{Frank}(u_1,u_2|\theta) = \theta (1-e^{-\theta})e^{-\theta (u_1+u_2)}[(1-e^{-\theta})-(1-e^{-\theta u_1})(1-e^{-\theta u_2})]^{-2},
\label{eq:FrankDensity}
\end{equation*}}\noindent
the density of the Joe copula with parameter $\theta \geq 1$ is
{\footnotesize
\begin{equation*}
\begin{split}
	c_{Joe}(u_1,u_2|\theta) &= \left((1-u_1)^{\theta} +(1-u_2)^{\theta}-(1-u_1)^{\theta}(1-u_2)^{\theta}\right)^{\frac{1}{\theta}-2} \cdot(1-u_1)^{\theta-1} \\
	&\cdot(1-u_2)^{\theta-1}\cdot [\theta-1+(1-u_1)^{\theta} +(1-u_2)^{\theta}-(1-u_1)^{\theta}(1-u_2)^{\theta}]	
\label{eq:JoeDensity}.
\end{split}
\end{equation*}}

\end{document}